\documentclass[12pt,epsf,qsymbols]{article}
\pdfoutput=1
\usepackage[utf8]{inputenc}
\usepackage[normalem]{ulem}
\usepackage[english]{babel}%,frenchb
\usepackage{tabularx}
\usepackage{array}
\usepackage{graphics}
\usepackage{graphicx}
\usepackage{psfrag}
\usepackage{epsfig}
\usepackage{amsmath}
\usepackage{amssymb}
\usepackage{bbold}
\usepackage{setspace}
\usepackage{rotating}
\usepackage{colortbl}
\usepackage{longtable}
\usepackage{slashed}
\usepackage{braket}
\usepackage{lineno}
\makeatletter
\usepackage{textcomp}
\usepackage[usenames,dvipsnames]{xcolor}
\usepackage{relsize}
\usepackage{cite}

\usepackage{bbm}
\usepackage{bm}
\usepackage{enumerate}
\usepackage{amsmath}
\usepackage{verbatim}

\usepackage{hyperref}
\hypersetup{colorlinks,citecolor=nicegreen,linkcolor=niceblue}
\hypersetup{colorlinks=true}

\setlongtables

\newcommand{\cb}{{\mathcal B}}
\newcommand{\bea}{\begin{eqnarray}}
\newcommand{\eea}{\end{eqnarray}}
\newcommand{\beq}{\begin{equation}}
\newcommand{\eeq}{\end{equation}}
\newcommand{\ec}{\end{center}}
\newcommand{\bc}{\begin{center}}

\newcommand{\tev}{{\rm TeV}}
\newcommand{\gev}{{\rm GeV}}

\newcommand{\pdir}{p\kern -5.2pt\raise 0.2ex\hbox {/}}

\newcommand{\vdir}{v\kern -5.75pt\raise 0.15ex\hbox {/}}
\newcommand{\kdir}{k\kern -5.75pt\raise 0.15ex\hbox {/}}
\newcommand{\epsdir}{\epsilon\kern -5.0pt\raise 0.15ex\hbox {/}}
\newcommand{\bvdir}{\bar{v}\kern -5.75pt\raise 0.15ex\hbox {/}}
\newcommand{\Ddir}{D\kern -7.75pt\raise 0.20ex\hbox {/}}
\newcommand{\Adir}{A\kern -7.75pt\raise 0.20ex\hbox {/}}
\newcommand{\ldir}{l\kern -5.0pt\raise 0.2ex\hbox{/}}
\newcommand{\varepsdir}{\varepsilon\kern -5.5pt\raise 0.15ex\hbox{/}}

\newcommand {\E}[1]{\times 10^{#1}}	% exponent notation
\newcommand{\mc}[1]{\mathcal{#1}}
\newcommand{\mrm}[1]{\mathrm{#1}}
\newcommand{\ellp}[0]{\ell^\prime}

\newcommand{\re}[0]{\mrm{Re}}

\makeatother

\definecolor{niceblue}{rgb}{0.15,0.15,0.6}
\definecolor{nicegreen}{rgb}{0.1,0.5,0.1}
\definecolor{Red}{rgb}{1.,0.,0.}

\definecolor{Green}{rgb}{0.2,.7,0.2}

\begin{document}
\unitlength = 1mm

\thispagestyle{empty} 
\begin{flushright}
\begin{tabular}{l}
{\tt \footnotesize LPT 16-53}\\
\end{tabular}
\end{flushright}
\begin{center}
\vskip 3.4cm\par
{\par\centering \textbf{\LARGE  
\Large \bf Palatable Leptoquark Scenarios for Lepton Flavor \\ 
\vskip .3cm 
\Large \bf Violation in Exclusive $b\to s\ell_1\ell_2$ modes}}\\
\vskip 1.2cm\par
{\scalebox{.85}{\par\centering \large  
\sc D.~Be\v{c}irevi\'c$^a$, N.~Ko\v{s}nik$^{b,c}$, O.~Sumensari$^{a,d}$ and R.~Zukanovich Funchal$^{d}$}
{\par\centering \vskip 0.7 cm\par}
{\sl 
$^a$~Laboratoire de Physique Th\'eorique (B\^at.~210)\\
CNRS and Univ. Paris-Sud, Universit\'e Paris-Saclay, 91405 Orsay cedex, France.}\\
{\par\centering \vskip 0.25 cm\par}
{\sl 
$^b$~Departement of Physics, University of Ljubljana\\ Jadranska 19, 1000 Ljubljana, Slovenia.}\\
{\par\centering \vskip 0.25 cm\par}
{\sl 
$^c$~Jo\v{z}ef Stefan Institute\\ Jamova 39, P.O. Box 3000, 1001 Ljubljana, Slovenia.}\\
{\par\centering \vskip 0.25 cm\par}
{\sl 
$^d$~Instituto de F\'isica, Universidade de S\~ao Paulo, \\
 C.P. 66.318, 05315-970 S\~ao Paulo, Brazil.}\\

{\vskip 1.65cm\par}}
\end{center}

\vskip 0.85cm
\begin{abstract}
We examine various scenarios that involve a light ${\cal O}(1\ \tev)$
leptoquark state and select those which are compatible with the
current experimental values for $\mathcal{B}(B_s\to \mu\mu)$,
$\mathcal{B}(B\to K\mu\mu)_{\mathrm{large}-q^2}$,
$R_K=\mathcal{B^\prime}(B\to K\mu\mu)/\mathcal{B^\prime}(B\to Kee)$,
and which lead to predictions consistent with other experimental
data. We show that two such scenarios are phenomenologically
plausible, namely  the one with a doublet of scalar leptoquarks of
hypercharge $1/6$, and the one with a triplet of vector leptoquarks of
hypercharge $2/3$. We also argue that a model with a singlet scalar
leptoquark of hypercharge $1/3$ is not viable.  Using the present
experimental data as constraints, it is shown that the exclusive lepton
flavor violating decays, $\mathcal{B}(B_s\to \mu\tau)$,
$\mathcal{B}(B\to K \mu\tau)$ and $\mathcal{B}(B\to K^\ast \mu\tau)$,
can be as large as $\mathcal{O}(10^{-5})$.

\end{abstract}
\newpage
\setcounter{page}{1}
\setcounter{footnote}{0}
\setcounter{equation}{0}
%%%%%%%%%%%%%%%%%%%%%%%%%%%%%%%%%%%%%%%%
\noindent

\renewcommand{\thefootnote}{\arabic{footnote}}
%\linenumbers

\setcounter{footnote}{0}

\tableofcontents

\newpage

%%%%%%%%%%%
%%%%%%%%%%%
%%%%%%%%%%%
\section{Introduction}
\label{sec:intro}

The experimental searches of New Physics (NP) at the LHC are of great
importance in our quest for solving the hierarchy and flavor
problems. Rare decays and low-energy precision measurements play a
crucial role in searching for the effects of physics beyond the
Standard Model (BSM) because they are complementary to 
direct searches and even allow us to probe energy scales
well above those that can be probed through them. Among the various types of rare decays, the lepton flavor
violating (LFV) ones are particularly appealing because they are
absent in the Standard Model (SM), and their detection would represent
a clean signal of NP.

Theoretically, the LFV decays are predicted in various NP scenarios,
such as in models with heavy sterile
fermions~\cite{SterilesLFV}, in models with a peculiar
breaking of supersymmetry~\cite{SUSY}, in models involving an
extra $Z^\prime$-boson~\cite{Zprime}, and in multiple
scenarios founded on the paradigm of existence of  one or more 
leptoquark (LQ) states~\cite{Dorsner:2016wpm}. Most of the current experimental
efforts in searching for the effects of LFV are
focused onto lepton decays, $\ell_1\to\ell_2 \gamma$,
$\ell_1\to3\ell_2$ (with $\ell_{1,2}\in\{e,\mu,\tau\}$), and on the $\mu-e$ conversion in nuclei~\cite{REVIEW}. 
The experimental
sensitivity to probe these decays is expected to be improved in the
years to come by several orders of magnitude. The LFV decays of
hadrons are complementary to the purely leptonic decays because they
represent a convenient ground to test LQ models, and they also
represent a different environment to test the result of the
above-mentioned leptonic modes. For example, in the LQ models, which
will be discussed in this paper, the rates of the LFV decays of
hadrons are much larger than in radiative and three-body lepton decays.

In this paper we will focus on the exclusive LFV modes,
$B_s\to\ell_1\ell_2$ and $B\to K^{(\ast)}\ell_1\ell_2$. At present,
the most constraining experimental bound is
$\mathcal{B}(B_s\to e^\pm \mu^\mp)^{\mathrm{exp}}<1.1\times 10^{-8}$, which has been
recently improved by an order of magnitude~\cite{Aaij:2013cby}. The
only dedicated experimental searches with a $\tau$-lepton in the final
state have been made by the BaBar Collaboration which reported the
upper limits 
$\mathcal{B}(B^+\to K^+ e^\pm \tau^\mp)^{\mathrm{exp}}<3.0\times 10^{-5}$ and
$\mathcal{B}(B^+\to K^+ \mu^\pm \tau^\mp)^{\mathrm{exp}}<4.8\times
10^{-5}$~\cite{Lees:2012zz}.
The modes with one $\tau$-lepton in the final state are
phenomenologically more appealing because the relevant couplings are
less constrained by direct experimental limits and for that
reason, in the following, we will focus on $B_s\to\mu\tau$ and
$B\to K^{(\ast)}\mu\tau$.

One of the main motivations to study transitions based on $b\to s\ell_1\ell_2$ comes from the recent observation of lepton flavor universality (LFU) violation 
in~\footnote{Notice that, for shortness, we use $\cb^\prime (B\to K\ell\ell)$ for the partial branching fraction corresponding to $q^2\in [1,6]\, \gev^2$. }
\begin{align}
\label{eq:rk}
R_K =\frac{\mathcal{B^\prime}(B^+\to K^+ \mu \mu)}{\mathcal{B^\prime}(B^+\to K^+ e e)}\equiv  \left.\frac{\mathcal{B}(B^+\to K^+ \mu \mu)}{\mathcal{B}(B^+\to K^+ e e)}\right\vert_{q^2 \in [1,6]\,\mathrm{GeV}^2},
\end{align}
where the squared dilepton mass is integrated in the bin $q^2\in [1,6]$ $\mathrm{GeV}^2$. LHCb found~\cite{Aaij:2014ora}
\begin{equation}\label{eq:RKexp}
R_K^{\mathrm{exp}}=0.745^{+0.090}_{-0.074}(\mathrm{stat})\pm 0.036(\mathrm{syst}),
\end{equation}
which is 2.4$\sigma$ lower than the SM prediction $R_K^\mathrm{SM}=1.00(1)$ \cite{Bobeth:2007dw,Bordone:2016gaq}. Since the hadronic uncertainties largely cancel in the ratio, if confirmed, this result would be an unambiguous manifestation of NP. 
Another experimental observation of LFU violation has been made in the
processes mediated by the charged-current interaction, namely, 
\begin{equation}
R_{D^{(\ast)}} = \dfrac{\mathcal{B}(B\to D^{(\ast)} \tau \nu)}{\mathcal{B}(B\to D^{(\ast)} \ell \nu)}, \qquad \ell\in\{ e,\mu\}\,,
\end{equation}
which turned out to be $2\sigma$ (for $R_D$) and $3.4\sigma$ (for
$R_{D^\ast}$) larger than predicted by the
SM~\cite{Lees:2013uzd,Huschle:2015rga,Aaij:2015yra,Fajfer:2012vx,Becirevic:2012jf}. 
The possibility of connecting the observed LFU violation
in neutral- and charged- current processes triggered 
many interesting theoretical works, mostly based on introducing
various LQ states or new gauge bosons ($Z^\prime$ and
$W^\prime$)~\cite{RKpapers,Becirevic:2015asa,Fajfer:2015ycq,Bauer:2015knc,Greljo:2015mma}. Remarkably,
in most of the proposed scenarios, LFV in meson decays is
  induced at observable rates~\cite{Glashow:2014iga,Crivellin:2015era}. In a previous work, 
we computed the general expressions for the full angular distribution
of $B_s\to\ell_1\ell_2$ and $B\to K^{(\ast)}\ell_1\ell_2$
decay modes, and explored a generic $Z^\prime$ model, as well as the
possibility of a Higgs induced LFV~\cite{Becirevic:2016zri}. In this paper we make a phenomenological 
analysis of the exclusive $b\to s\ell_1\ell_2$ decays in the LQ models which are compatible with the observed $R_{K}$.

The remainder of this paper is organized as follows. In
Sec.~\ref{sec:eff} we remind the reader of the effective approach to
$b\to s\ell_1\ell_2$ decays and briefly summarize the findings of
Ref.~\cite{Becirevic:2016zri}. In Sec.~\ref{sec:models} we catalogue
the plausible LQ models which are then scrutinized in
Sec.~\ref{sec:pred} where we select the models compatible with the
available $b\to s \mu\mu$ data. The models
selected in that way, are then subjected to a phenomenological analysis in order to obtain
upper bounds on the LFV $b\to s\ell_1\ell_2$ exclusive decay modes presented in
Sec.~\ref{sec:predbis}. We summarize and conclude in
Sec.~\ref{sec:conc}. Most of the auxiliary formulas used in this paper
are collected in the Appendices.

%%%%%%%%%%%
%%%%%%%%%%% 
%%%%%%%%%%%
\section{Effective Approach}
\label{sec:eff}

The most general dimension-six effective Hamiltonian describing the
LFV transitions $b\to s\ell_1^- \ell_2^{+}$, with $\ell_{1,2}\in\{ e,\mu,\tau\}$) is defined by~\cite{Altmannshofer:2008dz}
\begin{equation}
\label{eq:hamiltonian}
\begin{split}
  \mathcal{H}_{\mathrm{eff}} = -\frac{4
    G_F}{\sqrt{2}}V_{tb}V_{ts}^* &\Bigg{\lbrace} \sum_{i=1}^6
  C_i(\mu)\mathcal{O}_i(\mu)+\sum_{i=7,8}
  \Big{[}C_i(\mu)\mathcal{O}_i(\mu)+\left(C_{i}(\mu)\right)^\prime \left(\mathcal{O}_{i}(\mu)\right)^\prime\Big{]}\\
& + \sum_{i=9,10,S,P}
  \Big{[} C^{\ell_1 \ell_2}_i(\mu)\mathcal{O}^{\ell_1 \ell_2}_i(\mu) + \left(C^{\ell_1 \ell_2}_{i}(\mu)\right)^\prime \left(\mathcal{O}^{\ell_1 \ell_2}_{i}(\mu)\right)^\prime\Big{]}\Bigg{\rbrace}
+\mathrm{h.c.},
\end{split}
\end{equation}
where $C_i(\mu)$ and $C_i^{\ell_1 \ell_2}(\mu)$ are the Wilson coefficients, while the effective operators relevant to our study are defined by
\begin{align}
\label{eq:C_LFV}
\begin{split}
\mathcal{O}_{9}^{\ell_1\ell_2}
  &=\frac{e^2}{(4\pi)^2}(\bar{s}\gamma_\mu P_{L}
    b)(\bar{\ell}_1\gamma^\mu\ell_2), \qquad\qquad\hspace*{0.4cm}
\mathcal{O}_{S}^{\ell_1\ell_2} =
  \frac{e^2}{(4\pi)^2}(\bar{s} P_{R} b)(\bar{\ell}_1 \ell_2),\\
    \mathcal{O}_{10}^{\ell_1\ell_2} &=
    \frac{e^2}{(4\pi)^2}(\bar{s}\gamma_\mu P_{L}
    b)(\bar{\ell}_1\gamma^\mu\gamma^5\ell_2),\qquad\qquad
\mathcal{O}_{P}^{\ell_1\ell_2} =
  \frac{e^2}{(4\pi)^2}(\bar{s} P_{R} b)(\bar{\ell}_1 \gamma^5 \ell_2),
\end{split}
\end{align}
in addition to the electromagnetic penguin operator,
$\mathcal{O}_7=e/(4\pi)^2m_b (\bar{s}\sigma_{\mu\nu}P_R
b)F^{\mu\nu}$. The chirality flipped operators $\mathcal{O}_i^\prime$ are obtained from $\mathcal{O}_i$ by the replacement $P_L\leftrightarrow P_R$. Using the above Hamiltonian, one can easily compute the amplitudes for $B_s\to\ell_1^-\ell^{+}_2$ and $B\to K^{(\ast)}\ell_1^-\ell^{+}_2$ decays, and derive the expressions for the corresponding decay rates~\cite{Becirevic:2016zri,Gratrex:2015hna}, also summarized in Appendix~\ref{app:brs} of the present paper. 
In the SM, the LFV Wilson coefficients, $C_i^{\ell_1\ell_2}$,  are
zero due to lepton flavor conservation. The inclusion of neutrino
masses can induce LFV at the loop-level, but with negligible rates due
to the smallness of the neutrino masses. 
In the following we shall assume that the only source of LFV is NP and
hence in order to compute the Wilson coefficients for $b\to s\ell_1\ell_2$ ($\ell_1 \neq\ell_2$) we need to specify the NP model.

To simplify our notation, in what follows, we will denote $C_i \equiv C_i^{\ell_1\ell_2}$  ($\ell_1 \neq\ell_2$) when confusion can be avoided. Notice, however, that: 
\begin{itemize}
\item[(i)] in general, for $\ell_1\neq\ell_2$, $C_i^{\ell_1\ell_2} \neq C_i^{\ell_2\ell_1}$, which is particularly the case in the LQ models in which LFV occurs through 
tree-level diagrams;
\item[(ii)] in some situations, even if 
 $C_i^{\ell_1\ell_2} = C_i^{\ell_2\ell_1}$, $\forall i$, one can still generate an asymmetry between LFV modes with different lepton charges, e.g. $\mathcal{B}(B_s\to \mu^-\tau^+)\neq\mathcal{B}(B_s\to \mu^+\tau^-)$, cf. expressions in Appendix \ref{app:brs}.
\end{itemize} 
Before discussing the issue of lepton charge asymmetry one
must first observe LFV, that is why we will here combine the two
charged modes, namely, 
$\mathcal{B}(B_s\to\ell_1\ell_2)\equiv\mathcal{B}(B_s\to
\ell_1^-\ell_2^+)+\mathcal{B}(B_s\to \ell_1^+\ell_2^-)$, and
$\mathcal{B}(B\to K^{(\ast)}\ell_1\ell_2)\equiv\mathcal{B}(B\to
K^{(\ast)} \ell_1^-\ell_2^+)+\mathcal{B}(B\to
K^{(\ast)}\ell_1^+\ell_2^-)$. 

We should also emphasize that the LFV channels we consider respect a hierarchy which depends on the adopted NP scenario. If the LFV is generated only by the (pseudo-)scalar operators, the lifted helicity suppression implies  
\begin{align}
C_{S,P}^{(\prime)}\neq 0, C_{9,10}^{(\prime)}=0:\hspace*{1.2cm}\mathcal{B}(B_s\to \ell_1\ell_2)>\mathcal{B}(B\to K \ell_1\ell_2)>\mathcal{B}(B\to K^\ast \ell_1\ell_2),
\end{align} 
whereas the inverted hiearchy is obtained if the nonzero Wilson coefficients are those corresponding to the (axial-)vector operators ($C_{S,P}^{(\prime)}= 0$, $C_{9,10}^{(\prime)}\neq 0$)~\cite{Becirevic:2016zri}.

%%%%%%%%%%%
%%%%%%%%%%% 
%%%%%%%%%%%
\section{$C_i^{\ell_1\ell_2}(\mu)$ in Leptoquark Models}
\label{sec:models}

Leptoquarks are colored states that can mediate
interactions between quarks and leptons. Such particles can appear in
Grand Unified Theories~\cite{Georgi:1974sy} and in models with
composite Higgs states~\cite{Schrempp:1984nj}, as recently reviewed
in~\cite{Dorsner:2016wpm}.  In general, a
  LQ can be a scalar or a vector field which in turn can come as an 
$SU(2)_L$-singlet, -doublet or -triplet~\cite{Buchmuller:1986zs}. In
the following we assume that the SM is extended by only one of the LQ
states and list the models which can be used to describe the
$b\to s \ell_1 \ell_2$ processes.  We adopt the notation of
Ref.~\cite{Kosnik:2012dj} and specify the LQ states by their quantum
numbers with respect to the SM gauge group, $(SU(3)_c, SU(2)_L)_Y$,
where the electric charge, $Q=Y+T_3$, is the sum of hypercharge ($Y$)
and the third weak isospin component ($T_3$).  Throughout this paper the flavor
indices of LQ couplings will refer to the mass eigenstates of
down-type quarks and charged leptons, unless stated otherwise. In
other words, the left-handed doublets are defined as
$Q_i = [(V^\dagger u_L)_i\; d_{Li}]^T$ and
$L_i = [(U\nu_L)_{i}\; \ell_{Li}]^T$, where $V$ and $U$ are the
Cabibbo-Kobayashi-Maskawa (CKM) and the
Pontecorvo-Maki-Nakagawa-Sakata (PMNS) matrices, respectively, $u_L$,
$d_L$, $\ell_L$ are the fermion mass eigenstates, whereas $\nu_L$
stand for the massless neutrino flavor eigenstates. Since the tiny
neutrino masses play no role for the purpose of this paper, we may
choose the PMNS matrix to be the unity matrix, $U=\mathbb{1}$.
Right-handed field indices will always refer to the mass eigenbasis.

\subsection{Scalar Leptoquarks}
\label{sec:slq}

We first consider three scalar LQ scenarios. Two of them can modify
the $b\to s\mu\tau$ decay at tree-level without spoiling the proton
stability.  A third scenario, which might destabilise the proton and will also be discussed below,
generates a contribution to $b\to s\mu\tau$ only at the loop-level.

\subsubsection{${ \Delta}^{(1/6)} \equiv (3,2)_{1/6}$}

This model has already been used to provide a viable explanation of $R_K^{\mathrm{exp}}$ in Ref.~\cite{Becirevic:2015asa}. The Yukawa Lagrangian reads, 
\begin{align}
\label{eq:slq2}
\begin{split}
\mathcal{L}_{\Delta^{(1/6)}} &= (g_L)_{ij} \bar{d}_{Ri} {\widetilde{\boldsymbol{\Delta}}}^{(1/6) \dagger} L_j+\mathrm{h.c.},\\
&=\left(g_L \right)_{ij}\bar{d}_i P_L
\nu_j\,\Delta^{(-1/3)}-(g_L)_{ij} \bar{d}_i P_L
\ell_j\,\Delta^{(2/3)}+\mathrm{h.c.},
\end{split}
\end{align}
where $\widetilde{\boldsymbol\Delta}\equiv i\tau_2 {\boldsymbol\Delta}^\ast$ is the conjugated
$SU(2)_L$ doublet, $g_L$ is a generic matrix of couplings. In the
second line we explicitly write the terms with 
$\Delta^{(-1/3)}$ and $\Delta^{(2/3)}$ where the superscripts refer to the electric
charge of the $Y=1/6$ LQ states. The masses of these two states can in
principle be different, but for simplicity we will assume them to be
equal, $m_{\Delta^{(2/3)}} = m_{\Delta^{(-1/3)}}\equiv
m_\Delta$.~\footnote{Note that the electroweak oblique corrections do
  not allow for large splittings between the two states of the doublet~\cite{Dorsner:2016wpm,Froggatt:1991qw,Keith:1997fv}.}. After integrating out the heavy fields, this model gives rise to the chirality flipped operators in the $b\to s\ell_1^-\ell_2^+$ effective Hamiltonian~(\ref{eq:hamiltonian}), and the corresponding Wilson coefficients are given by
\begin{equation}
\label{eq:sl2-wc}
\left(C_{9}^{\ell_1\ell_2}\right)^\prime = - \left(C_{10}^{\ell_1\ell_2}\right)^\prime  =- \frac{\pi v^2}{2 V_{tb}V_{ts}^\ast \alpha_\mathrm{em} }\frac{(g_L)_{s\ell_1} (g_L)_{b\ell_2}^\ast}{m_{\Delta}^2},
\end{equation}
where $v= 246$~GeV is the well known electroweak vacuum expectation value, $v^2 = 1/(\sqrt{2}G_F)$. 

\subsubsection{${\Delta}^{(7/6)} \equiv (3,2)_{7/6}$}

The relevant Lagrangian for this model reads,
\begin{align}
\label{eq:slq1}
\begin{split}
\mathcal{L}_{\Delta^{(7/6)}} &=(g_R)_{ij} \bar{Q}_i{\boldsymbol\Delta}^{(7/6)}\ell_{Rj}+\mathrm{h.c.},\\
&=(V g_R)_{ij}\bar{u}_i P_R \ell_j\,\Delta^{(5/3)}+(g_R)_{ij}\bar{d}_i P_R \ell_j\,\Delta^{(2/3)}+\mathrm{h.c.},
\end{split}
\end{align}
where $g_R$ denotes the Yukawa coupling matrix, while again 
the superscript in $\Delta^{(5/3)}$ and $\Delta^{(2/3)}$ refers to the electric charge of the two mass degenerate LQ states. 
After integrating out the heavy fields, the only non-zero Wilson coefficients relevant to $b\to s\ell_1^-\ell_2^+$ are,
\begin{equation}
\label{eq:sl1-wc}
C_9^{\ell_1\ell_2} = C_{10}^{\ell_1\ell_2}  = -\frac{\pi v^2}{2 V_{tb}V_{ts}^\ast \alpha_\mathrm{em}}\frac{(g_R)_{s\ell_1} (g_R)_{b\ell_2}^\ast}{m_{\Delta}^2},
\end{equation}
corresponding to the chirality non-flipped operators in Eq.~(\ref{eq:hamiltonian}).  

\subsubsection{$\Delta^{(1/3)} \equiv  (3,1)_{-1/3}$}

Being an electroweak singlet, this scalar LQ model is the simplest one. Its Lagrangian is given by,
\begin{align}
\label{eq:slq3}
\begin{split}
\mathcal{L}_{\Delta^{(1/3)}} &= (g_L)_{ij} \overline{Q_i^C} i \tau_2 L_j \Delta^{(1/3)\ast}+(g_R)_{ij} \overline{u_{Ri}^C} \ell_{Rj} {\Delta^{(1/3)\ast}} + \mathrm{h.c.}\\
&= {\Delta^{(1/3)\ast}}\Big[(V^\ast g_L)_{ij} \overline{u_{i}^C} P_L \ell_{j} -(g_L)_{ij}\overline{d_{i}^C} P_L \nu_j +(g_R)_{ij} \overline{u_{i}^C} P_R \ell_{j} \Big] + \mathrm{h.c.},
\end{split}
\end{align}
where the superscript $C$ stands for the charge conjugation.~\footnote{It should be clear that $g_{L,R}$ in Eq.~\eqref{eq:slq3} are entirely different couplings from those appearing in Eq.~(\ref{eq:slq2}) or in Eq.~(\ref{eq:slq1}).} 
In addition to terms shown in~\eqref{eq:slq3} one could also write terms involving diquarks. To
  avoid conflict with the proton decay bounds those couplings must be
  negligible, so we set them here to zero.

The Wilson coefficients contributing to $b\to s\ell_1^-\ell_2^+$ are generated at loop level and, to one-loop order, they are given by~\cite{Bauer:2015knc,Das:2016vkr}:
\begin{align}
\label{eq:sl3-wc}
C_9^{\ell_1\ell_2} - C_{10}^{\ell_1\ell_2} &= \frac{m_t^2}{8\pi \alpha_\mathrm{em} m_\Delta^2}(V^\ast g_L)_{t\ell_1}^\ast(V^\ast g_L)_{t\ell_2}-\frac{1}{64\pi\alpha_\mathrm{em}}\frac{v^2}{m_\Delta^2}\frac{(g_L\cdot g_L^\dagger)_{bs}}{V_{tb}V_{ts}^\ast}(g_L^\dagger \cdot g_L )_{\ell_1\ell_2},\\
\label{eq:sl3-wc-bis}
C_9^{\ell_1\ell_2} + C_{10}^{\ell_1\ell_2} &=  \frac{m_t^2}{16\pi \alpha_\mathrm{em} m_\Delta^2}(g_R)_{t\ell_1}^\ast(g_R)_{t\ell_2}\Bigg{[}\log \frac{m_\Delta^2}{m_t^2}-f(x_t)\Bigg{]}-\frac{1}{64\pi\alpha_\mathrm{em}}\frac{v^2}{m_\Delta^2}\frac{(g_L\cdot g_L^\dagger)_{bs}}{V_{tb}V_{ts}^\ast}(g_R^\dagger \cdot g_R )_{\ell_1\ell_2},
\end{align}
where $f(x_t)=1+\dfrac{3}{x_t-1}\left(\dfrac{\log x_t}{x_t-1}-1\right)$, and $x_t=m_t^2/m_W^2$.

\subsection{Vector Leptoquarks}
\label{sec:vlq}
Vector LQs naturally appear in Grand Unified Theories as gauge
bosons of the unified gauge group or in composite Higgs scenarios. Here we
shall focus on relatively light vector LQs that can contribute to LFV $b\to s \ell_1 \ell_2 $ transitions. 
There are three vector LQ states that could participate in $d_i \to d_j \ell_1 \ell_2$
processes via dimension-$4$ tree-level couplings to the SM fermions.
These states, $\boldsymbol U_3$, $\boldsymbol V_2$, and $U_1$, are respectively a 
weak triplet, doublet, and 
singlet under the $SU(2)_L$. Here we do not
consider the weak doublet vector $\boldsymbol V_2$, because its nonzero couplings might jeopardize the proton stability~\cite{Kosnik:2012dj}.

\subsubsection{$U_1 \equiv (3,1)_{2/3}$}
Being a $SU(2)_L$ singlet $U_1$ can 
couple to both the left-handed and right-handed fermions, namely,
\begin{align}
\label{eq:U1}
\begin{split}
\mc{L}_{U_1} &= x^{LL}_{ij} \bar Q_{i} \gamma_\mu  U_1^\mu L_j  + x^{RR}_{ij} \bar
d_{Ri} \gamma_\mu U_1^\mu \ell_{Rj} + \mrm{h.c.}\\
&= U_{1\mu}\Big[(V x^{LL})_{ij} \bar{u}_i \gamma^\mu P_L\nu_j +x^{LL}_{ij}d_i \gamma^\mu P_L\ell_j  + x^{RR}_{ij}\bar{d}_{i}\gamma^\mu P_R \ell_j\Big]+\mathrm{h.c.}, 
\end{split}
\end{align}
where $x^{LL}_{ij}$ and $x^{RR}_{ij}$ are the matrices of Yukawa
  couplings. The non-chiral nature of $U_1$ couplings to fermions induces both vector and scalar Wilson coefficients:
\begin{alignat}{2}
  \label{eq:U1Wilsons}
      C_9^{\ell_1 \ell_2}  &= -C_{10}^{\ell_1 \ell_2}  &&=
      -\frac{\pi v^2}{ V_{tb}V_{ts}^\ast\alpha_\mathrm{em} m_U^2} x^{LL}_{s\ell_2} x_{b
        \ell_1}^{LL\ast},\nonumber\\
      \left(C_{9}^{\ell_1 \ell_2}\right)^\prime  &= \left(C_{10}^{\ell_1 \ell_2} \right)^\prime &&=
      -\frac{\pi v^2}{ V_{tb}V_{ts}^\ast\alpha_\mathrm{em} m_U^2} x^{RR}_{s\ell_2} x_{b
        \ell_1}^{RR\ast},\\
      C_S^{\ell_1 \ell_2}  &= -C_{P}^{\ell_1 \ell_2}  &&=
      \frac{2\pi v^2}{ V_{tb}V_{ts}^\ast\alpha_\mathrm{em} m_U^2} x^{LL}_{s\ell_2} x_{b
        \ell_1}^{RR\ast},\nonumber\\
      \left(C_{S}^{\ell_1 \ell_2}\right)^\prime  &= \left(C_{P}^{\ell_1 \ell_2}\right)^\prime  &&=
      \frac{2\pi v^2}{ V_{tb}V_{ts}^\ast \alpha_\mathrm{em} m_U^2} x^{RR}_{s\ell_2} x_{b
        \ell_1}^{LL\ast},\nonumber
\end{alignat}
where $m_U$ is the LQ mass. Obviously this model provides a rich
ground for phenomenology because it gives rise to both the chirality
flipped and non-flipped operators.

\subsubsection{${ U_3} \equiv  (3,3)_{2/3}$}
The only possible dimension-$4$ gauge invariant interaction between $\boldsymbol U_3$, a weak
triplet of LQ states, and fermions reads 
\begin{equation}
\label{eq:U3}
\begin{split}
\mc{L}_{U_3} &= x^{LL}_{ij} \bar Q_{i} \gamma_\mu  \bm{\tau}\cdot \bm{U}_3^\mu
L_j + \mrm{h.c.}\\
&= U^{(2/3)}_{3\mu} \, \Big[ (V x^{LL})_{ij}\, \bar u_i \gamma^\mu P_L \nu_j  -
        x^{LL}_{ij} \,\bar d_i \gamma^\mu P_L \ell_j \Big]\\
 &\quad +U^{(5/3)}_{3\mu}\, (\sqrt{2} V x^{LL})_{ij}\, \bar u_i \gamma^\mu P_L
 \ell_j \\
  &\quad +U^{(-1/3)}_{3\mu}\, (\sqrt{2} x^{LL})_{ij}\, \bar d_i \gamma^\mu P_L
 \nu_j+\mrm{h.c.}
\end{split}
\end{equation}
where by $x^{LL}_{ij}$ we denote the couplings
to the SM quarks and lepton doublets.
The absence of diquark couplings guarantees that this LQ state 
is both baryon and lepton number conserving. 
The Wilson coefficients for the $b\to s \ell_1 \ell_2$ Hamiltonian are
in the case of $\boldsymbol U_3$ obtained by keeping nonzero only the $x^{LL}$ couplings in Eq.~(\ref{eq:U1Wilsons}). 
For completeness we write them explicitly, 
\begin{align}
      C_9^{\ell_1 \ell_2}  &= -C_{10}^{\ell_1 \ell_2}  =
      -\frac{\pi v^2}{ V_{tb}V_{ts}^\ast\alpha_\mathrm{em} m_U^2} x^{LL}_{s\ell_2} x_{b
        \ell_1}^{LL\ast},
\end{align}
where we assume the degeneracy among the charge eigenstates of the
triplet, $m_U\equiv m_{U^{(-1/3)}}=m_{U^{(2/3)}}=m_{U^{(5/3)}}$.

%%%%%%%%%%%
%%%%%%%%%%% 
%%%%%%%%%%%
\section{Confronting LQ models with data}
\label{sec:pred}

\subsection{Which leptoquark model?}
\label{sec:fit-bsmumu}

In this Section, from the list of models spelled out in
Sec.~\ref{sec:models}, we select those
that are consistent with the experimental data for the exclusive
$b\to s\mu\mu$ processes. To that end we require the consistency with
the measured
$\mathcal{B}(B_s\to\mu^+\mu^-)^{\mathrm{exp}}=(2.8^{+0.7}_{-0.6})\times
10^{-9}$~\cite{CMS:2014xfa},~\footnote{The
  quoted result was obtained after combining the CMS and LHCb results
  in Ref.~\cite{CMS:2014xfa}.  Since then the ATLAS Collaboration also
  measured this same decay mode and reported
  $\mathcal{B}(B_s\to\mu^+\mu^-)^{\mathrm{exp}}=(0.9^{+1.1}_{-0.8})\times
  10^{-9}$~\cite{Aaboud:2016ire}.
  Combining this result with the previous two would probably
  exacerbate the discrepancy with the SM but to do it properly one should
  combine the statistical samples of all three mentioned
  collaborations.  }  and
$\mathcal{B}(B\to K \mu^+\mu^-)_{ \text{high}-q^2}^{\mathrm{exp}}=(8.5\pm 0.3\pm 0.4 )\times10^{-8}$~\cite{Aaij:2014pli}, since in that region of 
$q^2\in [15,22]\, \gev^2$ the hadronic uncertainties are well controlled by means of
numerical simulations of QCD on the
lattice~\cite{Aoki:2016frl}. Furthermore, by choosing
$q^2 \in [15,22]\ \gev^2$ we also avoid the narrow charmonium
resonances so that we can rely on the quark-hadron
duality~\cite{Beylich:2011aq}. This is also the reason why we use only $\mathcal{B}(B_s\to\mu^+\mu^-)$ and $\mathcal{B}(B\to K\mu^+\mu^-)_{{\rm high} - q^2}$ to derive the constraints on possible NP contributions: (a) They involve the decay constant $f_{B_s}$ and a few form factors at large $q^2$'s, the quantities which have been accurately computed by means of lattice QCD; (b) These quantities are not plagued by uncertainties related to the $c\bar c$-resonances. 
Theoretical description of $B\to K^\ast\mu^+\mu^-$ decay entails more hadronic form factors and more assumptions are needed when confronting theory with experiment. Similarly, a recently observed discrepancy between theory and experiment regarding $\mathcal{B}(B_s\to\phi \mu^+\mu^-)$~\cite{Aaij:2015esa} relies on the assumption of validity of the light cone QCD sum rule estimates of the hadronic form factors at low and intermediate $q^2$'s for which the systematic uncertainties are hard to assess.  Furthermore, for the purpose of our study, two measured quantities are sufficient to bound the NP couplings and we choose those which require the least number of assumptions, i.e. $\mathcal{B}(B_s\to\mu^+\mu^-)$ and $\mathcal{B}(B\to K\mu^+\mu^-)_{{\rm high} - q^2}$. Notice, however, that our results are compatible, at a $(2\div 3)\sigma$ level, with the global fit analyses made in Ref.~\cite{global}.

\begin{itemize}
\item The ${\boldsymbol \Delta}^{(1/6)}$-model was studied in Ref.~\cite{Becirevic:2015asa} where it has been shown that the required consistency with the measured $\mathcal{B}(B_s\to\mu^+\mu^-)^{\mathrm{exp}}$ and $\mathcal{B}(B\to K \mu^+\mu^-)_{\text{high}-q^2}^{\mathrm{exp}}$ results in 
the $R_K$ values fully compatible with experiment.~\footnote{This model has also been considered in Ref.~\cite{Hiller:2014yaa} 
to explain $R_K^{\mathrm{exp}}$ but by enhancing $\mathcal{B}(B\to K ee)$, instead of decreasing $\mathcal{B}(B\to K \mu\mu)$.}  
In this scenario $(C_{9})^\prime=-(C_{{10}})^\prime$, and from the combined fit of $\mathcal{B}(B_s\to\mu^+\mu^-)$ and $\mathcal{B}(B\to K \mu^+\mu^-)_{\text{high}-q^2}$ we obtain to $2\sigma$ accuracy,
\bea
(C_{9}^{\mu\mu})^\prime \in (-0.41,-0.08)_{\scriptscriptstyle \rm HPQCD},\ (-0.48,-0.17)_{\scriptscriptstyle \rm MILC},
\eea
where the results in the first and second interval are obtained by using the hadronic form factors computed in lattice QCD in Ref.~\cite{Bouchard:2013pna} and Ref.~\cite{Bailey:2015dka}, respectively. Notice that the first interval coincides with what is obtained in 
Ref.~\cite{Becirevic:2015asa}. In what follows we will choose 
\begin{align}
\label{eq:c9pc10p}
(C_{9}^{\mu\mu})^\prime \in (-0.48,-0.08)\, , 
\end{align}
which covers both of the above intervals. 

\item For the  ${\boldsymbol\Delta}^{(7/6)}$-model, the NP Wilson coefficients satisfy the relation $C_{9}=C_{10}$. To scrutinize this scenario, we perform a fit of $C_{9,10}$ assuming them to be real and by applying the strategy described above. 
The allowed regions at $1\sigma$ and $2\sigma$ are
shown in Fig.~\ref{fig:c9c10} where we also draw the lines
corresponding to $C_9=\pm C_{10}$.  Clearly, the line $C_9=C_{10}$
does not touch the region allowed by
$\mathcal{B}(B_s\to\mu^+\mu^-)^{\mathrm{exp}}$ and
$\mathcal{B}(B\to K \mu^+\mu^-)_{{\rm high}-q^2}^{\mathrm{exp}}$ 
to $2\sigma$. For that reason we discard the scenario ${\boldsymbol\Delta}^{(7/6)}$ from further discussion.~\footnote{We should add, however, that the ${\boldsymbol \Delta}^{(7/6)}$ model remains a good candidate to explain $R_{D^{(\ast)}}^{\mathrm{exp}}$~\cite{Dorsner:2013tla} if the NP couplings to muons are negligible.} 

%%%%%%%%%%%%%%%%%%%%%%%%%%%%%%%%%%%%%%%%
%%%%%%%%%%%%%%%%%%%%%%%%%%%%%%%%%%%%%%%%
\begin{figure}[ht!]
\centering
\includegraphics[width=0.65\linewidth]{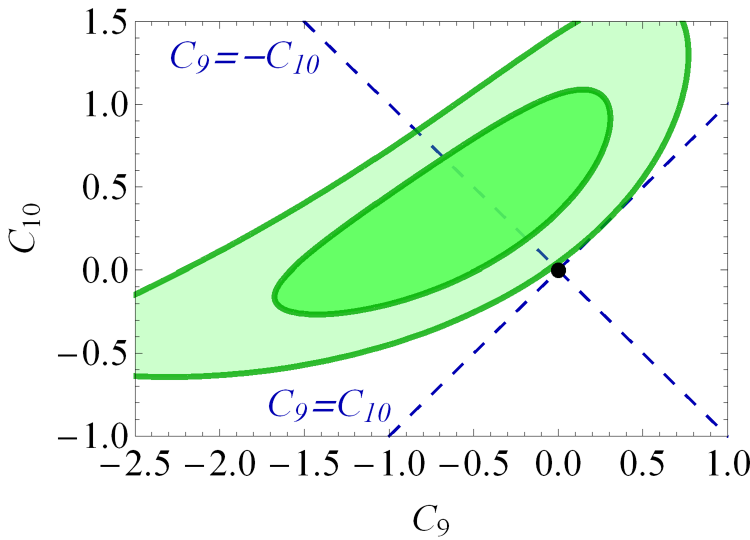}
\caption{\small \sl Regions in the plane
  $(C_9^{\mu\mu},C_{10}^{\mu\mu})$ that are in agreement with the
  experimental values of 
  $\cb(B_s\to\mu\mu)$ and $\cb(B^+\to K^+\mu\mu)_{{\rm large}\ q^2}$ at $1\sigma$ (dark green) and
$2\sigma$ (light green) accuracy. The black point represents the SM prediction. 
The dashed lines correspond to the scenarios $C_9^{\mu\mu}=-C_{10}^{\mu\mu}$ and $C_9^{\mu\mu}=C_{10}^{\mu\mu}$. }
\label{fig:c9c10}
\end{figure}
%%%%%%%%%%%%%%%%%%%%%%%%%%%%%%%%%%%%%%%%
%%%%%%%%%%%%%%%%%%%%%%%%%%%%%%%%%%%%%%%%
 
\item The $\Delta^{(1/3)}$-model generates both $C_9-C_{10}$ and $C_9+C_{10}$. In other words, this model is compatible 
with exclusive $b\to s\mu\mu$ data only if the combination with
the plus sign is suppressed. One can achieve that by taking $(g_R)_{ij}\approx 0$ in Eq.~(\ref{eq:sl3-wc-bis}), or by fine-tuning the terms in the same equation to get $C_9+C_{10}\approx 0$.
We will discuss more extensively this model in the next subsection. 

\item The ${\boldsymbol U_3}$-model was already studied in Ref.~\cite{Fajfer:2015ycq}
where it has been shown that it can provide a simultaneous explanation of
$R_K^{\mathrm{exp}}$ and $R_{D^{(\ast)}}^{\mathrm{exp}}$. In this scenario
$C_9=-C_{10}$, which is in agreement with the measured $\mathcal{B}(B_s\to\mu^+\mu^-)^{\mathrm{exp}}$ 
and $\mathcal{B}(B\to K \mu^+\mu^-)_{{\rm high}-q^2}^{\mathrm{exp}}$, as shown in Fig.~\ref{fig:c9c10}. 
To $2\sigma$ accuracy we find 
\bea
C_{9}^{\mu\mu}\in (-0.71,-0.04)_{\scriptscriptstyle \rm HPQCD},\ (-0.76,-0.11)_{\scriptscriptstyle \rm MILC}.
\eea
As before, we will use 
\begin{align}
\label{eq:c9c10}
C_{9}^{\mu\mu}\in (-0.76,-0.04) 
\end{align}
to cover the results obtained by using both sets of lattice QCD form factors.

\item A similar discussion applies to the $U_1$ LQ in the limit of 
$|x^{RR}_{ij}|\ll|x^{LL}_{ij}|$ in Eq.~(\ref{eq:U1Wilsons}). We should emphasize, however, that this model is not phenomenologically sound because the couplings to $\tau$'s are very poorly constrained by data. 
Since one cannot compute the loop corrections with vector LQs without specifying the ultraviolet (UV) 
completion or providing an explicit UV cut-off, no constraints from $B_s-\bar{B}_s$ and $\tau\to\mu\gamma$ can be used. 
While in the $\boldsymbol U_3$-model one can obtain a constraint by requiring the
consistency between the tree-level process $\mathcal{B}(B\to K
\nu\nu)$ and the upper experimental bound, in the $U_1$-model the
latter process is induced only at the loop-level 
and the experimental result is not constraining anymore. 
For this reason we prefer to discard the $U_1$-model
and focus only on the scenario with $\boldsymbol U_3$, as far as vector LQs are concerned. 

\end{itemize}

\

In summary, of all the LQ models discussed in  Sec.~\ref{sec:models},  we 
were able to discard in this section the scalar
${\boldsymbol \Delta}^{(7/6)}$ and the vector $U_1$. 
The remaining models can now be used to explore what are the allowed
$b\to s\mu\tau$ LFV decay rates. 
Before passing into that part, we  revisit the $\Delta^{(1/3)}$-model.

\subsection{On (in)viability of $\Delta^{(1/3)} \equiv (3,1)_{-1/3}$}
\label{sec:pred-1o3}
The interest in the $\Delta^{(1/3)}$-model has
been recently revived after the authors of Ref.~\cite{Bauer:2015knc}
claimed that it can simultaneously accommodate the observed LFU
violation in $R_K$ and $R_{D^{(\ast)}}$.  In this section we revisit
that claim and find that $R_K$ actually cannot be made significantly
smaller than one, without running into serious difficulties with other
measured quantities.

As we already mentioned above, 
the effective coefficients $C_{9}^{\ell_1\ell_2}\pm
C_{10}^{\ell_1\ell_2}$ are loop induced
and the results are given in Eqs.~(\ref{eq:sl3-wc})-(\ref{eq:sl3-wc-bis}). 
To suppress the combination with the plus sign, which is disfavored by the current $b\to s \mu\mu$ exclusive data, 
we set $g_R= 0$ in Eq.~(\ref{eq:slq3}).~\footnote{Another possibility
  is to impose the cancellation between the terms in
  Eq.~(\ref{eq:sl3-wc-bis}) but since the right-handed couplings lift the helicity suppression in leptonic decay rates one runs into problems with 
$\Gamma(K\to\mu\nu)$ and $\Gamma(D_{(s)}\to\mu\nu)$, and can also significantly enhance $\Gamma(\tau\to\mu\gamma)$, cf. Eq.~(\ref{eq:taumug1o3}) in Appendix~\ref{app:a1} of the present paper. In other words, the $g_R$ couplings are tightly constrained by data to be small, which justifies the approximation $g_R= 0$. } 
The left-handed couplings are assumed to have the following flavor structure
\begin{equation}
\label{eq:texture-slq-1o3}
g_{L} =
\begin{pmatrix}
  0 & 0 & 0\\
  0 & (g_{L})_{s \mu} & (g_{L})_{s \tau}\\
  0 & (g_{L})_{b \mu} & (g_{L})_{b \tau}
\end{pmatrix},
\qquad
V g_{L} =
\begin{pmatrix}
  0 & V_{us} (g_{L})_{s\mu} + V_{ub} (g_{L})_{b\mu} & V_{us} (g_{L})_{s\tau} + V_{ub} (g_{L})_{b\tau}\\
  0 & V_{cs} (g_{L})_{s\mu} + V_{cb} (g_{L})_{b\mu} & V_{cs} (g_{L})_{s\tau}  + V_{cb} (g_{L})_{b\tau}\\
  0 & V_{ts} (g_{L})_{s\mu} + V_{tb} (g_{L})_{b\mu} & V_{ts} (g_{L})_{s\tau}  + V_{tb} (g_{L})_{b\tau}
\end{pmatrix},
\end{equation}
where the first matrix connects down-type quarks to neutrinos, and the second  up-type quarks to charged leptons.~\footnote{We reiterate that the PMNS matrix $U$ is set to unity since the neutrino masses are neglected in this study.}
Notice that the couplings to the first generation of leptons are
assumed to be zero since they are tightly constrained by data. In our approach, the couplings to the $d$ quark are also set to
zero, since their nonzero values can induce  
significant contributions to processes such as
$K\to\pi\nu\nu$ and the $K-\bar{K}$ mixing. Even though we fix
$(g_L)_{d\mu}=(g_L)_{d\tau}=0$, from Eq.~(\ref{eq:texture-slq-1o3}),
we see that the couplings to the $u$ quark are generated via the CKM
matrix. Therefore, constraints from the kaon and the $D$-meson sectors
should be taken into account too. 
Besides those, one of the most important constraints comes from the
$B_s-\bar{B}_s$ mixing amplitude. We obtain
\begin{align}
\label{eq:Bsmix-D1o3}
\frac{\Delta m_{B_s}^{\mathrm{th}}}{\Delta m_{B_s}^{\mathrm{SM}}}=1+\frac{\eta_1 (g_L \cdot g_L^\dagger)^2_{bs}}{32 G_F^2 m_W^2 |V_{tb}V_{ts}^\ast|^2{\eta_B} S_0(x_t)m_\Delta^2},
\end{align}
where $\eta_1=0.82(1)$ accounts for the QCD running from $\mu=m_\Delta
\simeq 1$~TeV down to $\mu=m_b$, $S_0(x_t)$ is the Inami-Lim
  function, and $\eta_B$ encodes the short distance QCD corrections. 
We combine the experimental value $\Delta m_{B_s}^\mathrm{exp}=17.7(2)\;\mathrm{ps}^{-1}$~\cite{Agashe:2014kda}, with the SM prediction $\Delta m_{B_s}^{\mathrm{SM}}=17.3(17)\;\mathrm{ps}^{-1}$, to get $\Delta m_{B_s}^\mathrm{exp}/\Delta m_{B_s}^{\mathrm{SM}}=1.02(10)$ \cite{Becirevic:2016zri}. 
Furthermore, $R_{\nu\nu}=\mathcal{B}(B\to
K\nu\nu)_\mathrm{th}/\mathcal{B}(B\to K\nu\nu)_\mathrm{SM}$, for
which the expression will be given later on
[cf. Eq.~(\ref{eq:BKnunuformula})], should satisfy
$R_{\nu\nu}^{\mathrm{exp}}<4.3$, as 
established by BaBar~\cite{Lees:2013kla,Buras:2014fpa}. A pecularity of this LQ model is the modification of the (semi-)leptonic decays of pseudoscalar mesons. We define the effective Lagrangian of the transitions $d\to u\ell \bar{\nu}$ as
	\begin{align}
		\label{eq:lagrangian-lep-semilep}
		\mathcal{L}_{\mathrm{eff}} = -2\sqrt{2}G_F V_{u d}\Big{[}(1+g_V)(\overline{u}_{L}\gamma_\mu {d}_{L}) (\overline{\ell}_L\gamma^\mu\nu_L) &+g_S(\mu)(\overline{u}_{R} d_{L})(\overline{\ell}_R \nu_L)\\
		&+g_T(\mu)(\overline{u}_R \sigma_{\mu\nu}d_L)(\overline{\ell}_R \sigma^{\mu\nu} \nu_L)\Big{]}+\mathrm{h.c.},\nonumber
	\end{align}

\noindent where $u$ and $d$ stand for a generic up- and down-type quark flavor. 
Using this Lagrangian one can compute the (semi-)leptonic decay rates for the specific channels, e.g. $B\to D\ell\nu_{\ellp}$ and $B\to\ell\nu_{\ellp}$, as described in Appendix~\ref{app:a1}. After integrating out the LQ state, we obtain the effective coefficients
	\begin{align}
		\label{eq:semilep-gv}
		g_V\Big{\vert}_{d\to u\ell {\nu}_{\ellp}} &= \frac{1}{4\sqrt{2}G_F V_{ud}}\frac{(g_L)_{u\ell}^\ast (g_L)_{d\nu_{\ellp}}}{m_\Delta^2},\qquad\quad\\
		\label{eq:semilep-gs}
		g_S(\mu=m_\Delta)\Big{\vert}_{d\to u\ell {\nu}_{\ellp}}&=-\frac{1}{4\sqrt{2}G_F V_{ud}}\frac{(g_R)_{u\ell}^\ast (g_L)_{d\nu_{\ellp}}}{m_\Delta^2},\\
		\label{eq:semilep-gt}
		g_T(\mu=m_\Delta)\Big{\vert}_{d\to u\ell {\nu}_{\ellp}} &=-\frac{1}{4}g_S(\mu=m_\Delta)\Big{\vert}_{d\to u\ell {\nu}_{\ellp}},
	\end{align}
which can be inserted into the expressions for decay rates, explicitly given in Appendix~\ref{app:a1}, cf. Eqs.~(\ref{eq:lep})--(\ref{eq:semilep}). 
We should stress that LQs can induce new contributions in which the neutrino has a different flavor from the charged lepton. One should therefore sum over the unobserved neutrino flavors in order to compare with the experimentally measured rates, 
e.g. $\mathcal{B}(B\to D\ell\nu)=\sum_{\ellp}\mathcal{B}(B\to D\ell\nu_{\ellp})$, with $\ellp\in\{\mu,\tau\}$.

Considering the ansatz given in Eq.~(\ref{eq:texture-slq-1o3}) for the
Yukawa matrix, the relevant leptonic modes for our study are $K\to\mu{\nu}$,
$D_s\to (\mu,\tau){\nu}$, and $B\to \tau{\nu}$. We consider the
experimental values given in Ref.~\cite{Agashe:2014kda} and we use the
values for the decay constants computed in lattice QCD, summarized in
Appendix \ref{app:had}. Other observables, such as
$\mathcal{B}(\tau\to\mu\gamma)$ and $\mathcal{B}(D^0\to\mu\mu)$, for which the formulas are
given in Appendix~\ref{app:a1}, give important constraints as well.

\begin{figure}[h!]
\hspace*{-6mm}\includegraphics[width=0.52\linewidth]{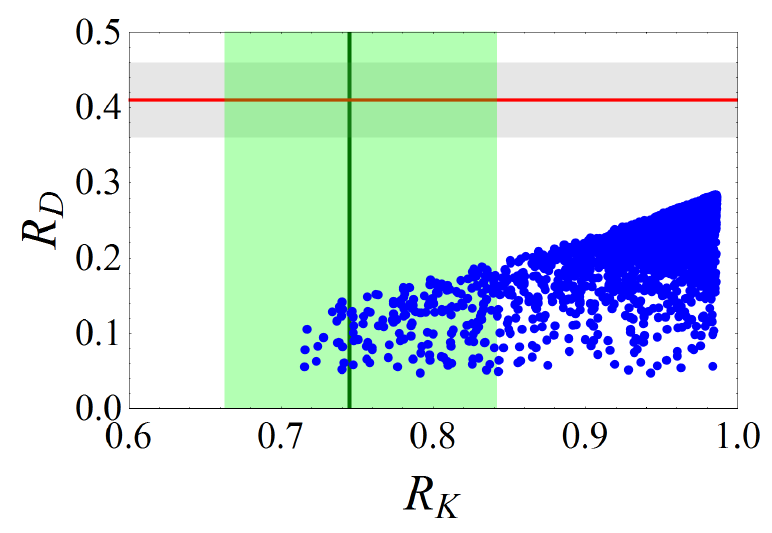}~\includegraphics[width=0.52\linewidth]{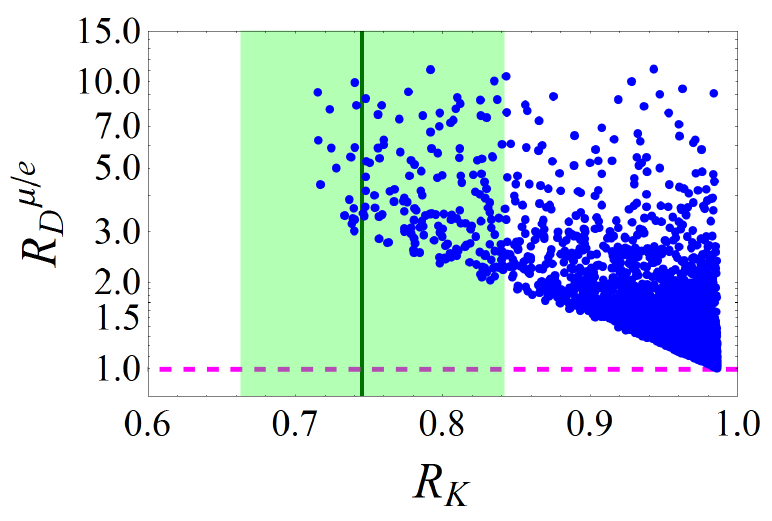}
\caption{\small \sl $R_D\equiv R_D^{\tau/\ell}$ (left panel) and $R_D^{\mu/e}$ (right panel) are plotted against $R_K$. The allowed points are compared with the experimental values (gray and green bands) at $1\sigma$. 
The dashed magenta line on the right panel corresponds
to the SM prediction for $R_D^{\mu/e}\simeq 0.995$. The theory error
bars for $R_D$ are not shown on the left panel. They are estimated to be of the order of a few percent.}
\label{fig:slq3-rkrd}
\end{figure}

In addition to the above-mentioned constraints, compactly collected in Tab.~\ref{tab:observables}, we impose the conditions of perturbativity, $|(g_L)_{ij}|<4\pi$, and look for the points which would simultaneously satisfy the observed $R_K^{\mathrm{exp}}$ and $R_D^{\mathrm{exp}}$. As it can be seen 
in Fig.~\ref{fig:slq3-rkrd} (left panel), we were not able to find couplings $(g_L)_{ij}$ that would result in values for $R_D$
consistent with experiment, $R_D^\mathrm{exp}=0.41(5)$
\cite{Lees:2013uzd,Huschle:2015rga}. In fact, the couplings of
$\Delta^{(1/3)}$ to the muon, which are necessary to get $R_K<1$, are
large enough and push $R_D$ to values smaller than the SM one.~\footnote{In obtaining $R_D$ we used the $B\to D$ form factors recently computed in lattice QCD~\cite{Lattice:2015rga}, which also give $R_D^{\rm SM}=0.286(12)$. }
Furthermore, even though we were able to find points that give acceptable values for $R_K$, we find that the selected points are in conflict with 
\begin{equation}
R_D^{\mu/e}=\dfrac{\mathcal{B}(B\to D\mu \nu)}{\mathcal{B}(B\to D e\nu)},
\end{equation}
as depicted in the right panel of Fig.~\ref{fig:slq3-rkrd}. Although
$R_D^{\mu/e}$ has not been experimentally established, values of $R_D^{\mu/e}\approx 1.05$ seem
already implausible, since the $\mathcal{B}(B\to D e \nu)$ and
$\mathcal{B}(B\to D\mu \nu)$ data have been successfully combined in $B$-factory experiments to extract ${\cal
  G}(1)|V_{cb}|$.
In Ref.~\cite{Greljo:2015mma} it was even argued that such a deviation
from lepton flavor universality cannot be larger than $2\%$. In
Fig.~\ref{fig:slq3-rkrd}, however, we see that the points selected to
satisfy the observed $R_K^{\mathrm{exp}}$ result in $R_D^{\mu/e}> 1.8$, 
a large departure from one. We, therefore, conclude that one cannot
accommodate the experimental value 
$R_K^{\mathrm{exp}}$ without producing a huge enhancement of
$\mathcal{B}(B\to D \mu \nu)$, which implies
$R_D \lesssim R_D^{\rm SM}$ and unacceptably large
$R_D^{\mu/e}$.~\footnote{In obtaining $R_D^{\mu/e}> 1.8$ we used
  $m_t^{\rm pole}$. Had we used $m_t^{\overline{\rm MS}}(m_t)$ we
  would have obtained $R_D^{\mu/e}> 1.5$, still much larger than
  $R_D^{\mu/e}\lesssim 1.05$.}  We were insisting on the agreement
with the experimental value $R_K^{\mathrm{exp}}$. If,
instead, one wants to accommodate the experimental value 
$R_D^{\mathrm{exp}}$ with this model~\cite{Dorsner:2013tla}, the resulting values of
the couplings to muon are either zero or far too small to explain
$R_K^{\mathrm{exp}}$.

%%%%%%%%%%%%%%%%%%%%%%%%%%%%%%%%%%%%%%%%
%%%%%%%%%%%%%%%%%%%%%%%%%%%%%%%%%%%%%%%%
%%%%%%%%%%%%%%%%%%%%%%%%%%%%%%%%%%%%%%%%
%%%%%%%%%%%%%%%%%%%%%%%%%%%%%%%%%%%%%%%%

There are several sources of disagreement between the present paper and Ref.~\cite{Bauer:2015knc}, which we comment on in detail. 
First of all, their best fit values for $R_{D^{(\ast)}}$ were obtained in a setup where NP only couples to $\tau$'s. This assumption is not justified, since a putative explanation of $R_K$ through loops necessarily 
requires large couplings to muons, which then modifies the denominator in $R_D$, and produces excessively large $R_D^{\mu/ e}$, as discussed above. 
Secondly, the conditions for the LQ parameters 
given by Eqs.~(12) and (17) of Ref.~\cite{Bauer:2015knc} are necessary but not sufficient to reproduce results consistent with $\mathcal{B}(B\to K\nu\nu)^{\mathrm{exp}}$ and $R_K^{\mathrm{exp}}$, respectively. Finally, the bounds from processes 
involving the up-quark, such as $K\to\mu{\nu}$, $D_s\to (\mu,\tau){\nu}$ and $B\to \tau{\nu}$, have been tacitly neglected in their work. As mentioned before, one cannot completely avoid the first quark generation, see Eq.~(\ref{eq:texture-slq-1o3}). 

\section{Results and discussion}
\label{sec:predbis}

In this section we present our predictions for the remaining viable scenarios, namely the LQ models ${\boldsymbol \Delta}^{(1/6)}$ and $\boldsymbol U_3$.
\begin{table}[ht!]
\centering
\begin{tabular}{|c|c|c|c|c|c|}
\hline 
Observable & Eqs. - ${\boldsymbol\Delta}^{(1/6)}$  &  Eqs. - $\Delta^{(1/3)}$ &  Eqs. - $\boldsymbol U_3$ & Exp. value  & Ref. \\ \hline\hline
$b\to s\mu\mu$	& (\ref{eq:sl1-wc}) & (\ref{eq:sl3-wc}) & (\ref{eq:U3Wilsons}) & Eqs.~(\ref{eq:c9pc10p})-(\ref{eq:c9c10})	& --	\\[0.25em] 
$\Delta m_{B_s}/\Delta m_{B_s}^\mathrm{SM}$ & (\ref{eq:Bsmix-D1o6}) &
                                                                      (\ref{eq:Bsmix-D1o3})
                                                             & --  & 1.02(10)	& \cite{Agashe:2014kda} \\[0.25em] 
$R_{\nu\nu}$ & (\ref{eq:BKnunuformula}) & (\ref{eq:BKnunuformula}) & \eqref{eq:U3_Rnunu}	& $<4.3$ &  \cite{Lees:2013kla,Buras:2014fpa} \\[0.25em] 
$\mathcal{B}(\tau\to\mu\gamma)$ & (\ref{eq:taumug1o6}) & (\ref{eq:taumug1o3}) & --& $<4.4 \times 10^{-8}$	&	\cite{Aubert:2009ag}\\[0.25em] 
$\mathcal{B}(\tau\to\mu\phi)$ & (\ref{eq:taumuphi}) & -- & \eqref{eq:taumuphiV} & $<8.4 \times 10^{-8}$	& \cite{Agashe:2014kda} \\[0.25em] 
$\mathcal{B}(D^0\to\mu^+\mu^-)$ & -- & (\ref{eq:D0mumu}) & \eqref{eq:U3-Dmumu} & $<7.6 \times 10^{-9}$ &  \cite{Aaij:2013cza} \\[0.25em]  
$\mathcal{B}(K\to\mu{\nu})$ & -- & (\ref{eq:semilep-gv})-(\ref{eq:semilep-gt}), (\ref{eq:lep}) & \eqref{eq:U3semilep},\eqref{eq:lep}  &  $63.58(11)\times 10^{-2}$	&	\cite{Rosner:2015wva} \\[0.25em] 
$\mathcal{B}(D_s\to\mu{\nu})$ & -- & (\ref{eq:semilep-gv})-(\ref{eq:semilep-gt}), (\ref{eq:lep}) &  \eqref{eq:U3semilep},\eqref{eq:lep}  & $5.56(25)\times 10^{-3}$	&  \cite{Agashe:2014kda} 	\\[0.25em] 
$\mathcal{B}(D_s\to\tau{\nu})$ &	-- & (\ref{eq:semilep-gv})-(\ref{eq:semilep-gt}), (\ref{eq:lep}) &	 \eqref{eq:U3semilep},\eqref{eq:lep} & $5.54(24)\times 10^{-2}$  &  \cite{Agashe:2014kda}   \\[0.25em]
$\mathcal{B}(B\to\tau{\nu})$ & -- & (\ref{eq:semilep-gv})-(\ref{eq:semilep-gt}), (\ref{eq:lep}) &  \eqref{eq:U3semilep},\eqref{eq:lep} & $1.44(32)\times 10^{-4}$	&	\cite{Lees:2012ju,Kronenbitter:2015kls} \\[0.25em]
$R_D$ &	 --  & (\ref{eq:semilep-gv})-(\ref{eq:semilep-gt}), (\ref{eq:semilep}) & \eqref{eq:U3semilep},\eqref{eq:semilep}& 0.41(5)	&  \cite{Lees:2013uzd,Huschle:2015rga}\\ \hline
\end{tabular}
\caption{\label{tab:observables} \small \sl Observables considered in
  our phenomenological analyses and their
  corresponding experimental values (or bounds) as well as their theoretical expressions for the scenarios ${\boldsymbol\Delta}^{(1/6)}$, $\Delta^{(1/3)}$ and $\boldsymbol U_3$.}
\end{table}

\subsection{${ \Delta}^{(1/6)} \equiv (3,2)_{1/6}$}
\label{sec:pred-1o6}

In our analysis of the ${\boldsymbol\Delta}^{(1/6)}$-model we assume the LQ couplings to the first generation of fermions to be negligible because they are most tightly constrained by the experimental limits on $\mu-e$
conversion on nuclei, on atomic parity violation, on
$\mathcal{B}(K_L\to\mu e)$, and on $\mathcal{B}(B_s\to \mu
e)$. Therefore, the matrix of couplings $(g_L)_{ij}$ in Eq.~(\ref{eq:slq2}) is assumed to have the form
\begin{equation}
\label{eq:couplingmatrix-1o6}
g_L=\begin{pmatrix}
0 & 0 & 0\\ 
0 & (g_L)_{s\mu} & (g_L)_{s\tau}\\ 
0 & (g_L)_{b\mu} & (g_L)_{b\tau}
\end{pmatrix},
\end{equation}
where the entries are considered to be real. 
The product $(g_L)_{s\mu}(g_L)_{b\mu}^\ast \neq 0$ is then fixed by Eqs.~(\ref{eq:sl2-wc}) and (\ref{eq:c9pc10p}). To constrain the couplings to $\tau$'s we use the information from $B_s-\bar{B_s}$ mixing amplitude, and the experimental limits on $\mathcal{B}(B\to K\nu\nu)$ and on $\mathcal{B}(\tau\to\mu\phi)$, see also Tab.~\ref{tab:observables}. To that end we recall the expression for the mass difference of the $B_s-\bar{B}_s$ system with respect to the SM one,
	\begin{align}
		\label{eq:Bsmix-D1o6}
		\frac{\Delta m_{B_s}^{\mathrm{th}}}{\Delta m_{B_s}^{\mathrm{SM}}}=1+\frac{\eta_1  (g_L \cdot g_L^\dagger)^2_{bs}}{16 G_F^2 m_W^2 |V_{tb}V_{ts}^\ast|^2\eta_B S_0(x_t)m_\Delta^2},
	\end{align}		
where $(g_L\cdot g_L^\dagger)$ denotes the product of matrices defined
in Eq.~(\ref{eq:couplingmatrix-1o6}).
%$S_0(x_t)$ is the Inami-Lim function, and $\eta_B$ encodes the short distance QCD corrections. We combine the experimental value $\Delta m_{B_s}^\mathrm{exp}=17.7(2)\;\mathrm{ps}^{-1}$~\cite{Agashe:2014kda}, with the SM prediction $\Delta m_{B_s}^{\mathrm{SM}}=17.3(17)\;\mathrm{ps}^{-1}$, to get $\Delta m_{B_s}^\mathrm{exp}/\Delta m_{B_s}^{SM}=1.02(10)$ \cite{Becirevic:2016zri}. 
The ratio $R_{\nu\nu}\equiv \mathcal{B}(B\to K\nu\nu)_\mathrm{th}/\mathcal{B}(B\to K\nu\nu)_\mathrm{SM}$, in this model, is modified as follows:
	\begin{align}
	\label{eq:BKnunuformula}
	R_{\nu\nu} =1 - \frac{1}{6\, C_L^{\mathrm{SM}}} \mathrm{Re}\left[ \dfrac{(g_L\cdot g_L^\dagger)_{sb}}{N  m_\Delta^2}\right]+ \frac{1}{48 (C_L^{\mathrm{SM}})^2 }\dfrac{(g_L \cdot g_L^\dagger)_{ss}(g_L \cdot g_L^\dagger)_{bb}}{ |N|^2 m_\Delta^4 },
	\end{align}
where $N = {G_F V_{tb} V_{ts}^\ast
  \alpha_\mathrm{em}}/{(\sqrt{2}\pi)}$ in the case of $\Delta^{(1/6)}$ and $N = -{G_F V_{tb} V_{ts}^\ast
  \alpha_\mathrm{em}}/{(\sqrt{2}\pi)}$ in the case of $\Delta^{(1/3)}$. The SM contribution is contained in 
$C_L^{\mathrm{SM}}=-6.38(6)$ as defined in
Ref.~\cite{Altmannshofer:2009ma}. 
%With the current experimental bound, $\mathcal{B}(B^+\to K^+\nu\bar{\nu})<1.7 \times 10^{-5}$ at 90\%CL \cite{Lees:2013kla}, one then has $R_{\nu\nu}<4.3$~\cite{Buras:2014fpa}. 
Finally, the experimental bound $\mathcal{B}(\tau\to\mu\phi)^{\mathrm{exp}}<8.4\times 10^{-8}$~\cite{Agashe:2014kda} implies the relation
\begin{equation}
\frac{|(g_L)_{s\tau} (g_L)_{s\mu}^\ast|}{m_\Delta^2}< 0.036\, \ \mathrm{TeV}^{-2}\quad (90\%\,\mathrm{CL}),
\end{equation}
which was derived using the expression given in Appendix~\ref{app:a1}. 

Other observables, such as leptonic decays of $\phi(nS)$ and $\Upsilon(nS)$ mesons, could in principle provide useful bonds on the LQ couplings but the derived limits turn out to be much less significant at this point. 
Flavor conserving leptonic decays, such as $\Upsilon(nS)\to\tau\tau$, are dominated by the very large tree-level electromagnetic contribution which undermines their sensitivity to NP. LFV modes, such as $\Upsilon\to\mu\tau$, are in principle sensitive to the LQ contribution but the current 
experimental limit $\mathcal{B}(\Upsilon(1S)\to\mu\tau)^{\mathrm{exp}}<6\times 10^{-6}$~\cite{Love:2008ys} is still too weak to be useful. 
We have also checked that in this model the upper experimental limit
on $\mathcal{B}(\tau\to \mu \gamma)$ does not provide any additional
constraint because of the cancellation of terms $\propto 1/m_\Delta^2$ in the loop function,
see Eq.~(\ref{eq:taumug1o6}) in Appendix~\ref{app:a1}.

\begin{table}[ht!]
\centering
\begin{tabular}{|c|c|c|c|}
\hline 
Quantity & $m_\Delta=1$ TeV & $m_\Delta=5$ TeV & $m_\Delta=10$ TeV\\ \hline\hline
$\mathcal{B}(B_s\to\mu\tau)$ & $< 1.0 \times 10^{-5}$  &  $<3.0 \times 10^{-6}$ & $<1.8 \times 10^{-7}$ \\  
$\mathcal{B}(B\to K\mu\tau)$ & $< 1.1 \times 10^{-5}$ & $<3.4 \times 10^{-6}$  & $<2.0 \times 10^{-7}$ \\  
$\mathcal{B}(B\to K^\ast\mu\tau)$ & $< 2.0 \times 10^{-5}$  &  $<6.1\times 10^{-6}$ & $< 3.7 \times 10^{-7}$ \\
  \hline
\end{tabular}
\caption{\small \sl Predictions for exclusive $B_{(s)}$ meson decays
  at 90\% CL for the ${\boldsymbol \Delta}^{(1/6)}$-model.}
\label{tab:boundsBR1o6} 
\end{table}

Using the constraints discussed above, and summarized in
Tab.~\ref{tab:observables}, in addition to the perturbativity
prerequisite, which we here decide to be $|(g_L)_{ij}|\leq 1$,  
we were able to scan the parameter space and find points
which are consistent with all the requirements. With the points
selected in that way, we then compute the Wilson coefficients
$(C_9^{\mu\tau})^\prime = - (C_{10}^{\mu\tau})^\prime$ by using
Eq.~(\ref{eq:sl2-wc}), and then insert them into
Eqs.~(\ref{expr1}),(\ref{expr2}) and (\ref{expr3}) to compute the
branching fractions.  The resulting values (upper bounds) are listed
in Tab.~\ref{tab:boundsBR1o6}.  Notice that the ratios of the
exclusive LFV modes in this model are independent of the Wilson
coefficients and given by
\begin{equation}\label{eq:ratioS}
\dfrac{\mathcal{B}(B\to K^\ast\mu\tau)}{\mathcal{B}(B\to K\mu\tau)}\approx 1.8 \qquad\quad\mathrm{and}\qquad\quad \dfrac{\mathcal{B}(B_s\to \mu\tau)}{\mathcal{B}(B\to K\mu\tau)}\approx 0.9,
\end{equation}
where in evaluating the above ratios we used the central values of the hadronic parameters. 
% Clearly the resulting bounds depend on the LQ mass and they are larger for smaller $m_\Delta$, as it can be noted in Tab.~\ref{tab:boundsBR1o6}.
Since most flavor observables are studied at tree-level, our predictions depend only on the ratios $(g_L)_{d^\prime \ell}/m_\Delta$ with $d^\prime=d,s,b$ and $\ell=\mu,\tau$. The only exception is $\Delta m_{B_s}$, 
which exhibits a mild dependence on $m_\Delta$. We see from
the results presented in Tab.~\ref{tab:boundsBR1o6} that the bounds
become tighter for larger $m_\Delta$ which is a consequence of the condition $|(g_L)_{ij}|\leq 1$.

%For that reason, the mass dependence of our predictions only becomes prominent in the large mass region, where the perturbativity condition is important. Therefore, the decay rates are expected to be smaller for large LQ masses as can be seen in Tab.~\ref{tab:boundsBR1o6}. \footnote{Notice that our predictions depend on the choice of the perturbativity condition, which can be relaxed if one considers $|g_L|<\sqrt{4 \pi}$ instead of $|g_L|<1$.}% We have checked that our predictions are robust for masses around the TeV scale.

 %%%%%%%%%%%%%%%%%%%%%%%%%%%%%%%%%%%%%%%%
%%%%%%%%%%%%%%%%%%%%%%%%%%%%%%%%%%%%%%%%
\begin{figure}[h!]
\centering
\includegraphics[width=0.65\linewidth]{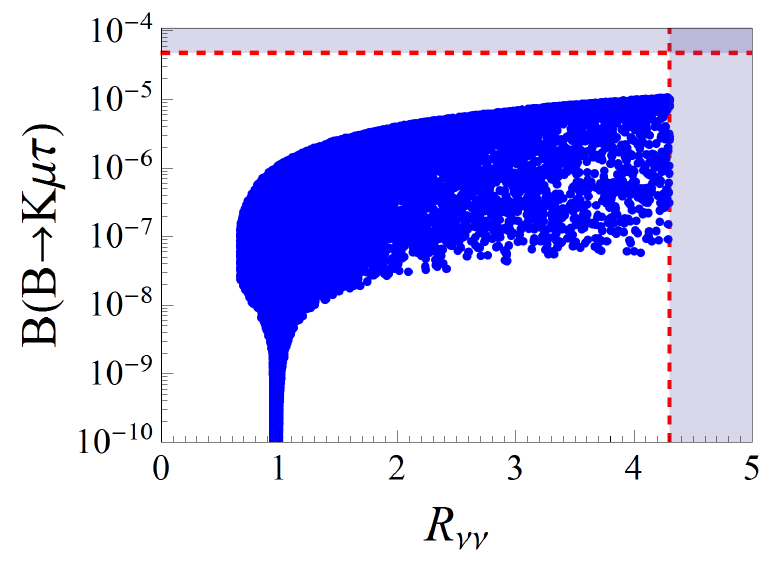}
\caption{\small {\sl $\mathcal{B}(B\to K\mu\tau)$ is plotted against
    $R_{\nu\nu}=\mathcal{B}(B\to
    K\nu\nu)_{\mathrm{th}}/\mathcal{B}(B\to
    K\nu\nu)_{\mathrm{SM}}$ for the 
    ${\boldsymbol\Delta}^{(1/6)}$-model. The blue points are allowed by
    flavor observables, and the red dashed lines correspond to the
    current experimental bounds  
$R_{\nu\nu}^{\mathrm{exp}}<4.3$} \cite{Buras:2014fpa} 
and {$\mathcal{B}(B^+\to K^+ \mu^\pm \tau^\mp)^{\mathrm{exp}}<4.8\times
10^{-5}$~\cite{Lees:2012zz}}.}
\label{fig:slq1-corr1}
\end{figure}
%%%%%%%%%%%%%%%%%%%%%%%%%%%%%%%%%%%%%%%%
%%%%%%%%%%%%%%%%%%%%%%%%%%%%%%%%%%%%%%%%

We next check on the correlation between $\mathcal{B}(B\to K\mu \tau)$ and $R_{\nu\nu}$ for a benchmark $m_\Delta=1$~TeV. We see from Fig.~\ref{fig:slq1-corr1} that the compatibility of $R_{\nu\nu}$ with the SM value, $R^{\rm SM}_{\nu\nu}=1$, 
would not exclude the possibility of rather large branching
fractions for the LFV decay modes. However, if the value of
$R_{\nu\nu}$ is found to be significantly below or above $1$ then this
analysis provides both an upper
and a lower bound for
$\mc{B}(B\to K\tau \mu)$. For example, if $R_{\nu\nu}\approx 2$, we get 
\begin{equation}
 2 \times 10^{-8} \lesssim \mathcal{B}(B\to K \mu\tau)\lesssim  5 \times 10^{-6}.
\end{equation}
Furthermore, we see that the current experimental bound, $R_{\nu\nu}^{\mathrm{exp}}\leq 4.3$, is an efficient constraint but it ceases to be so if we consider $m_\Delta \gtrsim 3$~TeV. 
Finally, it is interesting to note that the BaBar bound, $\mathcal{B}(B\to K \mu \tau)^{\mathrm{exp}}<4.8\times 10^{-5}$, is only an order of magnitude larger than the upper bound we obtained 
in our analysis. We hope that the high statistics LHCb data, and
especially those of the future Belle~II, will be used to improve this
as well as other bounds on similar decay modes. 
We reiterate that a bound on any of the modes considered above would be equally useful, since they are all related to each other through the ratios given in Eq.~(\ref{eq:ratioS}). 
Before closing this part of our analysis, we should mention that this LQ model has been extended to accommodate $R_D^{\rm exp}$ in Ref.~\cite{Becirevic:2016yqi}, without changing the upper bounds on the LFV rates discussed in this paper but 
providing the lower ones which are $\mathcal{O}(10^{-10})$. 

\subsubsection{Comment on $B_s\to\tau\tau$}

\begin{figure}[h!]
\centering
\includegraphics[width=0.65\linewidth]{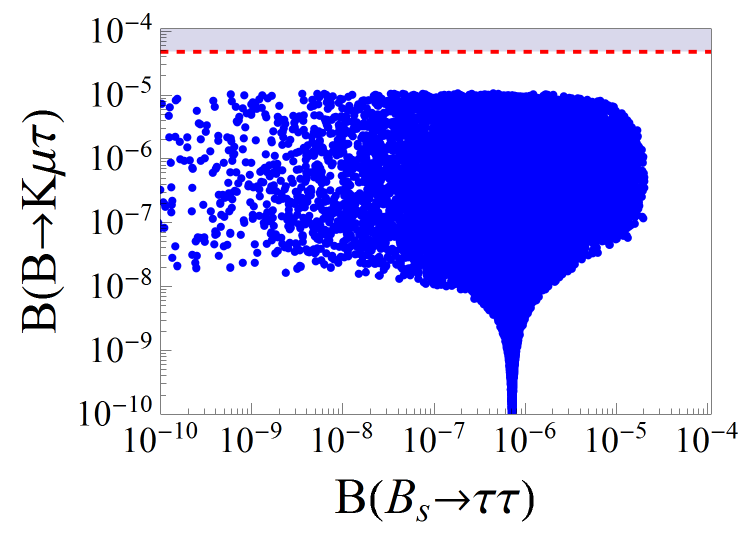}
\caption{\small \sl $\mathcal{B}(B\to K\mu\tau)$ is plotted against $\mathcal{B}(B_s\to\tau\tau)$ for 
the ${\boldsymbol\Delta}^{(1/6)}$-model. See Fig.~\ref{fig:slq1-corr1} for details.}
\label{fig:slq1-corr2}
\end{figure}

Since the LHCb Collaboration is actively searching for the $B_s\to\tau\tau$ events, we also show the correlation between $\mathcal{B}(B\to K\mu \tau)$ and $\mathcal{B}(B_s\to \tau\tau)$ in Fig.~\ref{fig:slq1-corr2}. 
We see again that even if $\mathcal{B}(B_s\to \tau\tau)$ is found to be consistent with the SM value, $\mathcal{B}(B_s\to \tau\tau)^{\rm SM}=7.7(5)\times 10^{-7}$, one can still have a significant  $\mathcal{B}(B\to K\mu \tau)$. 
This can be understood from Eq.~(\ref{eq:sl2-wc}), since $B_s\to \tau\tau$ will be modified by NP only if the product
$(g_L)_{s\tau}(g_L)_{b\tau}^\ast$ is nonzero, and therefore one should have both $(g_L)_{s\tau}\neq 0$ and $(g_L)_{b\tau}\neq 0$ to produce a deviation from the SM. 
On the other hand, the product $(g_L)_{s\mu}(g_L)_{b\mu}^\ast$ is
already fixed to a nonzero value
by $b\to s\mu\mu$ exclusive data. Therefore, it is enough to have one
of the couplings $(g_L)_{s\tau}$ or $(g_L)_{b\tau}$ different from
zero in order to induce LFV in $B_{(s)}$ decays through the effective
coefficients $(C_{10}^{\tau\mu})^\prime$ or
$(C_{10}^{\mu\tau})^\prime$, respectively. For that reason, the LFV
modes are more robust probes of the LQ couplings. Notice, however,
that if LHCb can establish an upper bound on
$\mathcal{B}(B_s\to\tau\tau)$, this would certainly be more important
than the other limits that we have used here to constraint the third generation couplings.

If, instead, the measured $\mathcal{B}(B_s\to \tau\tau)$ turns out to
be larger or smaller than $\mathcal{B}(B_s\to \tau\tau)^{\rm SM}$ this
model offers also a lower bound on  $\mathcal{B}(B\to K\mu \tau)$, which is quite remarkable because in this way one can 
check the validity of this scenario.  Moreover, we
see also that if 
$\mathcal{B}(B_s\to \tau\tau)\gtrsim 2\times 10^{-5}$ is measured then
this model not only cannot generate LFV, but it will be ruled out altogether.

%%%%%%%%%%%%%%%%%%%%%%%%%%%%%%%%%%%%%%%%
%%%%%%%%%%%%%%%%%%%%%%%%%%%%%%%%%%%%%%%%
%%%%%%%%%%%%%%%%%%%%%%%%%%%%%%%%%%%%%%%%
%%%%%%%%%%%%%%%%%%%%%%%%%%%%%%%%%%%%%%%%

\subsection{${ U_3} \equiv  (3,3)_{2/3}$} 

In a UV-complete renormalizable model containing $\boldsymbol U_3$ at the lower
end of the spectrum (and possibly accompanying light degrees of
freedom), we expect the $3\times 3$ matrix of LQ couplings $x^{LL}$ in
Eq.~\eqref{eq:U3} to be unitary. Indeed, if $\boldsymbol U_3$ was a remnant of a
larger multiplet of gauge bosons, one could consider a theory in the
unitary gauge in which the UV divergences appearing in the flavour changing neutral processes are removed by the sums over
the couplings, i.e. by the GIM mechanism, and for this technique to
work we would require $x^{LL}$ to be unitary.~\footnote{See also the
  discussion in Ref.~\cite{Biggio:2016wyy}.} However, in the unitary
case the couplings to $e$ and/or first generation of quarks cannot be
avoided and then the very strong null constraints from LFV searches
apply. In this particular case, we have found that the upper bounds on
$\mu-e$ conversion on Au nucleus and the one on
${\cal B}(K_L \to \mu e)$ already discard $\boldsymbol U_3$ with unitary $x^{LL}$
as a viable scenario for explaining $R_K^{\rm exp}$.

Thus we continue along with our analysis using the non-unitary ansatz,
and we completely avoid couplings to electrons by paying the price of being unable to unambiguously predict any
loop-induced process.  The ansatz below couples only the second and
third generation charged leptons with the down-type quarks. However,
as before, the CKM rotation will induce the couplings of $u,c,t$ to
the charged leptons and neutrinos, viz.
\begin{equation}
\label{eq:texture}
x^{LL} =
\begin{pmatrix}
  0 & 0 & 0\\
  0 & x^{LL}_{s \mu} & x^{LL}_{s \tau}\\
  0 & x^{LL}_{b \mu} & x^{LL}_{b \tau}
\end{pmatrix},
\qquad
V x^{LL} =
\begin{pmatrix}
  0 & V_{us} x^{LL}_{s\mu} + V_{ub} x^{LL}_{b\mu} & V_{us} x^{LL}_{s\tau} + V_{ub} x^{LL}_{b\tau}\\
  0 & V_{cs} x^{LL}_{s\mu} + V_{cb} x^{LL}_{b\mu} & V_{cs}  x^{LL}_{s\tau}  + V_{cb} x^{LL}_{b\tau}\\
  0 & V_{ts} x^{LL}_{s\mu} + V_{tb} x^{LL}_{b\mu} & V_{ts} x^{LL}_{s\tau}  + V_{tb} x^{LL}_{b\tau}
\end{pmatrix}.
\end{equation}

Observables that constrain this LQ have already been discussed in Ref.~\cite{Fajfer:2015ycq}.
Left-handed couplings of $U_{3\mu}^{(2/3)}$ alone allow for a single
combination of the Wilson coefficients:
  \begin{equation}
    \label{eq:U3Wilsons}
      C_9^{\ell_1 \ell_2}  = -C_{10}^{\ell_1 \ell_2}  =
      -\frac{\pi v^2}{V_{tb}V_{ts}^\ast \alpha_\mathrm{em} m_U^2} x^{LL}_{s\ell_2} x_{b
        \ell_1}^{LL\ast}.
\end{equation}
As discussed in Sec.~\ref{sec:fit-bsmumu}, by using the measured $\mathcal{B}(B_s\to\mu^+\mu^-)^{\rm exp}$  and $\mathcal{B}(B\to K \mu^+\mu^-)_{\text{high}-q^2}^{\rm exp}$, we select the 
values of $C_9^{\mu\mu}= -C_{10}^{\mu\mu}$, to $2\sigma$ accuracy and obtain $C_9^{\mu\mu} \in (-0.76, -0.04)$, cf. Eq.~(\ref{eq:c9c10}).
It is interesting to note that $\mathcal{B}(B_s\to\mu^+\mu^-)^{\rm exp}$ and $R_K^{\rm exp}$ impose almost degenerate constraints on $C_9^{\mu\mu}$. 
For example,  the result for $C_9^{\mu\mu}  = -C_{10}^{\mu\mu}$ to
$1\sigma$, results in  $R_K \in (0.74, 0.88)$, which fully agrees with
$R_K^{\rm exp}$.

%%%%%%%%%%%%%%%%%%%
The three charge components of $\boldsymbol U_3$, with the same
underlying couplings, allow for flavor effects both in the down/up-quarks
and the charged lepton/neutrino sectors. Interestingly, the
$2/3$-charge component %also
plays an important role in the charged current semileptonic
processes and thus can be used to address the
hint of LFU violation in $R_{D^{(*)}}$. Using the effective Lagrangian parametrization
of the charged current~\eqref{eq:lagrangian-lep-semilep} the
left-handed current modification reads
\begin{equation}
\label{eq:U3semilep}
g_V  \big{\vert}_{b\to c\ell {\nu}_{\ellp}} = -\frac{1}{2\sqrt{2} G_F V_{cb}} \frac{x^{LL*}_{b\ell}
  (V x^{LL})_{c \ellp}}{m_U^2}.
\end{equation}
Through $g_V$ the $\boldsymbol U_3$ vector LQ state modifies the overall
normalization of the SM spectrum of $B \to D \ell \bar\nu$ without
modifying its shape, cf. Eq.~\eqref{eq:semilep}
in Appendix~\ref{app:formulas}.

Thus in the $\boldsymbol U_3$-model we vary the couplings within
$|x^{LL}_{ij}| < 4\pi$ and apply the
above-mentioned constraints, namely the one on $C_{9}^{\mu\mu}$, and
the one arising from $R_D^\mrm{exp}$, both at $2\sigma$.~\footnote{To
  explain 
$R_D^{\mathrm{exp}}$ in this model,
  $|x_{b\mu}^{LL}|$ should be $\gtrsim 1$, which is why we opted for
  $|x^{LL}_{ij}| < 4\pi$, as often used in the literature as a
  perturbativity requirement for the couplings.}  We fix the mass at
$m_U = 1$~TeV. Further constraints are coming from the purely third-generation
charged current and the up-type neutral current processes which are mediated by the $U_{3\mu}^{(5/3)}$ eigencharge
component. Among the charged current processes, the most efficient
constraint comes from the measured ${\cal B}(t \to b \tau
\nu)^{\mathrm{exp}}= 0.096(28)$~\cite{Aaltonen:2014hua}. The
  effect of $\boldsymbol U_3$ here is an overall rescaling of $t\to b
  \tau \nu$ via coupling $g_V|_{b\to t \tau \nu_{\ell'}}$, cf. Eq.~\eqref{eq:U3semilep}.
We apply this constraint by eliminating
any point in the parameter space that results in ${\cal B}(t \to b \tau \nu)$ out of the  $1\sigma$ region of ${\cal B}(t \to b \tau \nu)^{\mathrm{exp}}$. 
Once the above constraints are applied it appears that the constraints coming from the 
up-quark rare process, $\mathcal{B}(D^0 \to \mu \mu)^{\mathrm{exp}}<
6.2\E{-9}$~\cite{Agashe:2014kda} (cf. Eq.~\eqref{eq:U3-Dmumu}), and from  
${\cal B}(\tau \to \mu \phi)$ are redundant.
Furthermore, we have checked that the LFU ratios, $\Gamma(\tau\to K\nu)/\Gamma(K\to \mu\nu)$ and  $\Gamma(K\to \mu\nu)/\Gamma(K\to e\nu)$, are consistent with experiment.

\begin{figure}[!t]
\hspace*{-9mm}\begin{tabular}{lr}
  \includegraphics[scale=0.65]{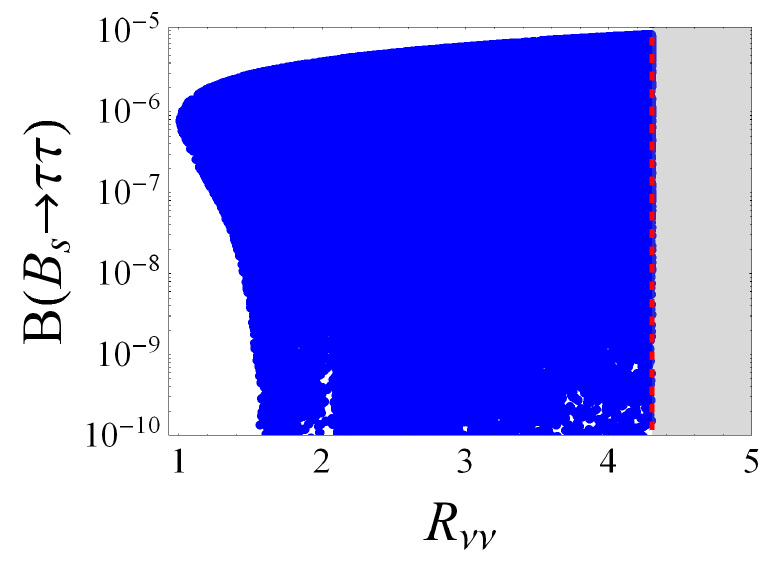} &
\includegraphics[scale=0.65]{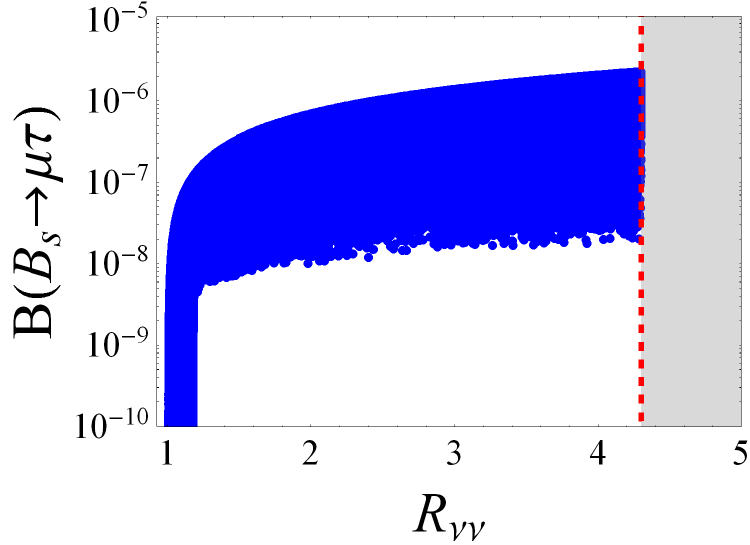}
\end{tabular}
  \caption{\small \sl Correlation between $R_{\nu\nu}$ and
    ${\cal B}(B_s \to \tau \tau)$ in the $\boldsymbol U_3$-model ($m_U=1$~TeV). The shaded grey region is excluded at 90\% CL by
the upper experimental bound on $R_{\nu\nu}$.}
  \label{fig:U3-nunu-tautau}
\end{figure}

Like in the previous subsection, with the couplings selected to verify the above constraints, we can compute various quantities.  
In particular, in Fig.~\ref{fig:U3-nunu-tautau} we show the correlation between
$R_{\nu\nu}$ and ${\cal B}(B_s \to \tau\tau)$, and again observe that the
current experimental limit on $R_{\nu\nu}$ provides a stringent
constraint on the parameter space. Note that $B \to K \nu\nu$ is
governed by the $Q=-1/3$ component which connects the down-type quarks
to neutrinos.  

Finally, on the right panel of Fig.~\ref{fig:U3-nunu-tautau} we show 
${\cal B}(B_s\to \mu\tau)$ with respect to $R_{\nu\nu}$, and we see
that this model allows for large LFV decay branching fractions
even if $R_{\nu\nu}$ and ${\cal B}(B_s \to \tau \tau)$ are close
  to their SM values. We obtain 
\bea
{\cal B}(B_s\to \mu\tau) \lesssim 2.7\times 10^{-6},
\eea 
for ${m_U}\simeq 1$~TeV. The bounds on the other two LFV modes can be obtained by simply using Eq.~(\ref{eq:ratioS}), i.e. 
\bea
\mathcal{B}(B\to K\mu\tau)\lesssim 3\times 10^{-6}, \qquad \mathcal{B}(B\to K^\ast\mu\tau)\lesssim 5.4\times 10^{-6}.
\eea 
The strong correlation between LFV and bi-neutrino modes is a signature
 of this model, should positive LFV signal or $R_{\nu\nu} > 1$ emerge.
 On the other hand, for the $B_s \to \tau\tau$ mode the left-hand panel in
 Fig.~\ref{fig:U3-nunu-tautau} demonstrates that it can be larger by
 about an order of magnitude than its SM prediction. For $R_{\nu\nu}> 1$ there is a possibility of strong
 suppression of $\mathcal{B}(B_s \to \tau \tau)$. Finally, the observation of
 $R_{\nu\nu}$ significantly smaller than $1$ is not compatible with
 the presence of $\boldsymbol U_3$ as an explanation for
 $R_K^{\mathrm{exp}}$ and $R_{D}^{\mathrm{exp}}$.  We should emphasise 
   that all the observables considered in this section have identical scaling with
   $\boldsymbol U_3$ couplings and mass, thus the correlations presented in Fig.~\ref{fig:U3-nunu-tautau}
   are kept intact for different choices of $m_U$, with the exception
   of perturbativity constraints which are becoming more stringent for
   increasing $m_U$.

\section{Summary}
\label{sec:conc}
%%%%%%%%%%%%%%%%%%%%%%%%%%%%%%%%%%
%%%%%%%%%%%%%%%%%%%%%%%%%%%%%%%%%%
In this paper we discuss exclusive $b\to s\ell_1\ell_2$ LFV
decay modes in various models in which the existence of an
$\mathcal{O}(1\ \tev)$ leptoquark state is assumed.  In doing so we
use the effective field theory approach and derive the expressions for the
Wilson coefficients in each of the considered scenarios, which then
allows us to test them against the available experimental
data for $\mathcal{B}(B_s\to\mu^+\mu^-)^{\rm exp}$ and
$\mathcal{B}(B\to K\mu^+\mu^-)^{\rm exp}$ at large $q^2$, and select
those which lead to predictions of other processes that are consistent
with current experimental data. These two quantities are used mostly
because the theoretical uncertainties are
under better control since the hadronic matrix elements have been estimated by means of numerical simulations
of QCD on the lattice, and also because the high-$q^2$ region is above
the very narrow $c\bar c$-resonances, so that the contribution of
$c\bar c$ can be treated by using the quark-hadron duality.

We find that two models (scenarios), compatible with the above-mentioned experimental constraints, 
can accommodate the experimental hint on LFU violation, namely  
$R_K^{\rm exp} = \mathcal{B^\prime}(B^+\to K^+ \mu \mu)/\mathcal{B^\prime}(B^+\to K^+ e e) < R_K^{\rm SM}$, 
and that they lead to the results in other processes that are compatible with experiment 
(and with the measured $\Delta m_{B_s}^{\mathrm{exp}}$, in
particular). Moreover, we find that the current experimental bound 
$R_{\nu\nu}^{\mathrm{exp}}$ is particularly useful 
in cutting a significant fraction of the parameter space. The two scenarios that we select as plausible are: 
\begin{itemize}
\item Scenario with ${\boldsymbol\Delta}^{(1/6)}$, a doublet of (mass degenerate)
  scalar leptoquarks of hypercharge $Y=1/6$, for which the NP
  contribution to $\mathcal{B}(B_s\to\mu^+\mu^-)$ and
  $\mathcal{B}(B\to K\mu^+\mu^-)$ comes from the left-handed 
  couplings, and for which the bound
  $\mathcal{B}(\tau\to \phi\mu)^{\rm exp}$ appears to be
  useful. Interestingly, we find that the LFV mode
  $\mathcal{B}(B\to K\mu\tau)$ can go up to $\mathcal{O}(10^{-6})$
  even if $R_{\nu\nu}=1$ and
  $\mathcal{B}(B_s\to\tau\tau)^{\rm exp}=
  \mathcal{B}(B_s\to\tau\tau)^{\rm SM}$,
  but that it is strictly different from zero for $R_{\nu\nu}\neq 1$
  and
  $\mathcal{B}(B_s\to\tau\tau)^{\rm exp}\neq
  \mathcal{B}(B_s\to\tau\tau)^{\rm SM}$.
  For example, for $R_{\nu\nu}\approx 2$, in this model we get
  $\mathcal{B}(B\to K\mu\tau) \in (0.2, 50)\times 10^{-7}$. Notice
  that this model is experimentally verifiable: (i) it leads to
  $R_{K^\ast}>1$~\cite{Becirevic:2015asa}, and (ii) it does not allow
  $R_{\nu\nu}\lesssim 0.6$, nor
  $\mathcal{B}(B_s\to\tau\tau)\gtrsim 2\times 10^{-5}$.
\item Scenario with $\boldsymbol U_3$, a triplet of (mass degenerate) vector leptoquarks of hypercharge $Y=2/3$, for which the NP contribution to  $\mathcal{B}(B_s\to\mu^+\mu^-)$ and $\mathcal{B}(B\to K\mu^+\mu^-)$ comes from the left handed couplings, 
and for which the bound on $R_{\nu\nu}^{\rm exp}$ as well as $\mathcal{B}(t\to b\tau\nu)^{\rm exp}$ provide the crucial constraints on its Yukawa couplings. Like in the previous case, the branching ratio $\mathcal{B}(B\to K\mu\tau)$ can be between zero and $\mathcal{O}(10^{-6})$ even if 
$R_{\nu\nu}=1$ and $\mathcal{B}(B_s\to\tau\tau)^{\rm exp}= \mathcal{B}(B_s\to\tau\tau)^{\rm SM}$, but it is stricly non-zero for $R_{\nu\nu}>1$. This model too is experimentally verifiable: (i) it leads to $R_{K^\ast} = 0.86(12)$, thus $R_{K^\ast}<1$~\cite{Hiller:2014yaa}, and (ii) it can be ruled out if $R_{\nu\nu}<1$ and/or $\mathcal{B}(B_s\to\tau\tau) \gtrsim 10^{-5}$.
\end{itemize}
As for the branching fractions of the similar LFV decay modes, they are related to $\mathcal{B}(B\to K\mu\tau)$ via Eq.~(\ref{eq:ratioS}), as discussed in Ref.~\cite{Becirevic:2016zri}.
Notice that we also examined the model with a singlet scalar leptoquark of hypercharge $Y=-1/3$ and found that it is not phenomenologically viable, i.e. that it cannot accommodate $R_K<1$ without running into serious phenomenological difficulties 
with other measured processes.  

\vskip 1.4cm
\noindent 
{\bf \large Acknowledgments:} {\sl We would like to thank S.~Fajfer for discussions, and P.~Arnan for pointing out to us the subtlety related to the interference term in Eq.~(31) of the present paper. R.Z.F. thanks the Universit\'e Paris Sud for the kind hospitality, and CNPq and FAPESP for partial financial support. N.K. acknowledges support of the Slovenian Research Agency. This project has received funding from the European Union's Horizon 2020 research and innovation program under the Marie Sklodowska-Curie grant agreement No. 674896. }

\newpage

\begin{appendix}

\section{Formulas and hadronic quantities}

In this Appendix we collect the expressions and numerical values used
in our analyses.

\label{app:formulas}

\subsection{Scalar Leptoquarks\label{app:a1}}

\begin{itemize}
    \item  \underline{$\tau\to\mu \phi$}:
We use the experimental upper bound $\mathcal{B}(\tau\to\mu \phi)<8.4 \times 10^{-8}$ \cite{Agashe:2014kda}, and the expression we derived,
\begin{equation}
\label{eq:taumuphi}
\mathcal{B}(\tau\to\mu \phi)=\frac{f_\phi^2 m_\phi^4}{256 \pi m_\tau^3 \Gamma_\tau}\left|\frac{g_{s\tau}g_{s\mu}^\ast}{m_\Delta^2} \right|^2 \left[-1+\frac{(m_\mu^2+m_\tau^2)}{2 m_\phi^2}+\frac{(m_\mu^2-m_\tau^2)^2}{2 m_\phi^4}\right]\lambda^{1/2}(m_\phi^2,m_\tau^2,m_\mu^2),
\end{equation}
where the couplings $g_{s\ell}$ can be either $(g_L)_{s\ell}$ for the model ${\boldsymbol\Delta}^{(1/6)}=(3,2)_{1/6}$, or $(g_R)_{s\ell}$ if the model ${\boldsymbol\Delta}^{(7/6)}=(3,2)_{7/6}$ is considered.

\

	\item  \underline{$\tau\to\mu \gamma$}:
We use the upper bound $\mathcal{B}(\tau\to \mu \gamma)<4.4 \times 10^{-8}$ \cite{Aubert:2009ag}, and the expression 
\begin{equation}
\mathcal{B}(\tau\to \mu \gamma)=\frac{\alpha_{\mathrm{em}}(m_\tau^2-m_\mu^2)^3}{4 m_\tau^3 \Gamma_\tau} \left( |\sigma_L|^2 + |\sigma_R|^2 \right),
\end{equation}
where the coefficients $\sigma_{L(R)}$ for the models ${\boldsymbol\Delta}^{(1/6)}$, ${\boldsymbol\Delta}^{(7/6)}$ and $\Delta^{(1/3)}$, to leading order in $1/m_\Delta$, read~\cite{Dorsner:2016wpm,Lavoura:2003xp}
\begin{align}
\label{eq:taumug1o6}
\sigma_L^{(1/6)} &=0, \qquad\sigma_R^{(1/6)} = - i (g_L)_{b\mu} (g_L)_{b\tau}^\ast\frac{N_C m_b^2 m_\tau}{96 \pi^2 m_\Delta^4} \left[\frac{5}{2}+\log\left(\frac{m_b^2}{m_\Delta^2}\right)\right], \\
\label{eq:taumug7o6}
\sigma_R^{(7/6)}&=0,\qquad \sigma_L^{(7/6)} =  i (g_R)_{b\mu} (g_{R})_{b\tau}^\ast\frac{N_C m_\tau}{64 \pi^2 m_\Delta^2}\left\lbrace 1+\frac{4}{3}\frac{m_t^2}{m_\Delta^2}\left[1+\log\left(\frac{m_t^2}{m_\Delta^2}\right)\right]\right\rbrace,\\
\label{eq:taumug1o3}
\sigma_{L(R)}^{(1/3)} &= -i \frac{N_C m_\tau}{192 \pi^2 m_\Delta^2}\Bigg{\lbrace}(g_{R(L)})_{t\tau}(g_{R(L)})_{t\mu}^\ast-\frac{m_t}{m_\tau}(g_{L(R)})_{t\tau}(g_{R(L)})_{t\mu}^\ast \Bigg{[} 14+8\log \left(\frac{m_t^2}{m_\Delta^2}\right) \Bigg{]}\Bigg{\rbrace},
\end{align}
with $N_C=3$ and $(g_{L(R)})_{u\ell}=\sum_{d} V_{u d}^\ast (g_{L(R)})_{d\ell}$ in the last equation, where $u(d)$ stand for generic up(down)-type quark flavors.

\

	\item  \underline{$D^0 \to \mu^+\mu^-$}: We use the experimental limit $\mathcal{B}(D^0\to\mu^+\mu^-)<7.6 \times 10^{-9}$ ($95\%$ CL) \cite{Aaij:2013cza} and the expression~\cite{Bauer:2015knc}
\begin{align}
	\label{eq:D0mumu}
	\mathcal{B}(D^0 &\to\mu^+\mu^-)=\frac{f_D^2 m_D^3}{256 \pi m_\Delta^4 \Gamma_D} \left(\frac{m_D}{m_c}\right)^2 \beta_\mu \Bigg{[} \beta_\mu^2 \left\vert (g_L)_{c\mu} (g_R)_{u\mu}^{\ast}-(g_R)_{c\mu} (g_L)_{u\mu}^{\ast}\right\vert^2\\&+\left\vert (g_L)_{c\mu}(g_R)_{u\mu}^\ast+(g_R)_{c\mu}(g_L)_{u\mu}^\ast+\frac{2 m_\mu m_c}{m_D^2}\left[(g_L)_{c\mu}(g_L)_{u\mu}^\ast+(g_R)_{c\mu}(g_R)_{u\mu}^\ast\right]\right\vert^2 \nonumber\Bigg{]},
\end{align}
where $m_c=m_c(m_\Delta)$ is the running charm quark mass.

\

	\item  \underline{$B\to\tau\nu$}\, and \underline{$B\to D\tau\nu$}:
Using the effective Lagrangian defined in Eq.~(\ref{eq:lagrangian-lep-semilep}) one can compute the leptonic and semileptonic decay rates for the specific channels, e.g. for $B\to D\ell\nu_{\ellp}$ and $B\to\ell\nu_{\ellp}$. We obtain
\begin{align}
	\label{eq:lep}
	\Gamma(B\to\ell \nu_{\ellp}) &=  \frac{G_F^2 m_B |V_{ub}|^2 }{8\pi}f_B^2 m_\ell^2 \left(1-\frac{m_\ell^2}{m_B^2} \right)^2 \left\vert \delta_{\ell\ellp}+g_V-g_S(\mu)\frac{m_B^2}{m_\ell(m_c+m_b)}\right\vert^2,
\end{align}
and
\begin{align}
	\label{eq:semilep}
	\dfrac{\mathrm{d}\Gamma}{\mathrm{d}q^2}(B\to D\ell\nu_{\ellp}) = \frac{G_F^2|V_{cb}|^2 }{192\pi^3 m_B^3}&|f_+(q^2)|^2 \Bigg{\lbrace}|\delta_{\ell\ellp}+g_V|^2 c_+^\ell (q^2)+|g_T(\mu)|^2 c_T^\ell(q^2) \left\vert \frac{f_T(q^2)}{f_+(q^2)}\right\vert^2 \nonumber\\
	&+c_{TV}^\ell(q^2)\, \mathrm{Re}\left[ (\delta_{\ell\ellp}+g_V)g_T^\ast(\mu)\right] \frac{f_T(q^2,\mu)}{f_+(q^2)} \nonumber\\ 
	&+c_0^\ell(q^2) \left\vert \delta_{\ell\ellp}+g_V+g_S(\mu) \frac{q^2}{m_\ell (m_d-m_u)} \right\vert^2 \left\vert\frac{f_0(q^2)}{f_+(q^2)}\right\vert^2 \Bigg{\rbrace},
\end{align}
where we emphasize that the NP contribution can allow for a different neutrino flavor $\ell\neq\ellp$ in the final state. The $B\to D$ form factors and the $f_B$ decay constant are defined in the Appendix \ref{app:brs}, and the phase-space functions $c_i^\ell(q^2)$ are given by \cite{Becirevic:2012jf}
\begin{align}
	c_+^\ell(q^2) &= \lambda^{3/2}(\sqrt{q^2},m_B,m_D)\left[ 1-\frac{3}{2}\frac{m_\ell^2}{q^2}+\frac{1}{2}\left(\frac{m_\ell^2}{q^2}\right)^3\right], \\
	c_0^\ell (q^2) &= m_\ell^2 \,\lambda^{1/2}(\sqrt{q^2},m_B,m_D)\dfrac{3}{2}\dfrac{m_B^4}{q^2} \left( 1-\dfrac{m_\ell^2}{q^2}\right)^2 \left( 1-\dfrac{m_D^2}{m_B^2}\right)^2, \\
	c_T^\ell(q^2) &=\lambda^{3/2}(\sqrt{q^2},m_B,m_D)\frac{8 q^2}{(m_B+m_D)^2}\left[1-3\left(\frac{m_\ell^2}{q^2}\right)^2+2\left(\frac{m_\ell^2}{q^2}\right)^3\right],\\
	c_{TV}^\ell(q^2)&=\frac{12 m_\ell}{m_B+m_D}\lambda^{3/2}(\sqrt{q^2},m_B,m_D)\left(1-\frac{m_\ell^2}{q^2} \right)^2,
\end{align}

\noindent with $\lambda(a,b,c)=[a^2-(b+c)^2] [a^2-(b-c)^2]$.
\end{itemize}

\subsection{Vector leptoquark $ U_3$}
\label{sec:app_U3}
\begin{itemize}
\item   \underline{$B \to K\nu\nu$}: The relative modification of the SM decay rate of $B \to K \nu \nu$ in
the presence of the $U_{3\mu}^{(-1/3)}$ reads~\cite{Dorsner:2016wpm}
\begin{equation}
  \label{eq:U3_Rnunu}
  \begin{split}   
R_{\nu\nu} = 1 &- \frac{2}{3 C_L^{\rm SM}} \,  \re\left[{  (x^{LL}\cdot x^{LL \dag}) _{sb}  \over N m_U^2 }\right] + 
  \frac{1}{3 |C_L^\mrm{SM}|^2}  {  (x^{LL}\cdot x^{LL \dag}) _{ss}  (x^{LL}\cdot x^{LL \dag}) _{bb}  \over | N|^2 m_U^4},
  \end{split}
\end{equation}
where $C_L^\mrm{SM} = -6.38 \pm 0.06$~\cite{Altmannshofer:2009ma}. 
\item \underline{$D^0 \to \mu^+\mu^-$}: The
branching ratio of the rare decay $D^0 \to \mu\mu$ is sensitive to the
CKM-rotated matrix $x^{LL}$:
\begin{equation}
  \label{eq:U3-Dmumu}
  \Gamma_{D^0 \to \mu \mu} = \frac{f_D^2 m_D m_\mu^2}{8\pi}
  \sqrt{1-\frac{4m_\mu^2}{m_D^2}} \frac{\left| (Vx^{LL})_{u\mu} (Vx^{LL})_{c\mu}^*\right|^2}{m_U^4}.
\end{equation}
    \item  \underline{$\tau\to\mu \phi$}: The expression relevant to this LFV decay mode in the  $\boldsymbol U_3$-model reads, 
\begin{equation}
\label{eq:taumuphiV}
\mathcal{B}(\tau\to\mu \phi)=\frac{f_\phi^2 m_\phi^4}{64 \pi m_\tau^3
  \Gamma_\tau}\left|\frac{x^{LL}_{s\tau}x^{LL\ast}_{s\mu}}{m_U^2}
\right|^2 \left[-1+\frac{(m_\mu^2+m_\tau^2)}{2
    m_\phi^2}+\frac{(m_\mu^2-m_\tau^2)^2}{2
    m_\phi^4}\right]\lambda^{1/2}(m_\phi^2,m_\tau^2,m_\mu^2).
% \mathcal{B}(\tau\to\mu \phi)=\frac{f_\phi^2 m_\phi^4}{256 \pi m_\tau^3 \Gamma_\tau}\left|\frac{g_{s\tau}g_{s\mu}^\ast}{m_\Delta^2} \right|^2 \left[-1+\frac{(m_\mu^2+m_\tau^2)}{2 m_\phi^2}+\frac{(m_\mu^2-m_\tau^2)^2}{2 m_\phi^4}\right]\lambda^{1/2}(m_\phi^2,m_\tau^2,m_\mu^2),
\end{equation}
\end{itemize}

\subsection{Exclusive $b\to s$ decay rates}
\label{app:brs}
We recall in this section the full decay rate expressions for the modes $B_s\to \ell_1\ell_2$ and $B\to K^{(\ast)}\ell_1\ell_2$ \cite{Becirevic:2016zri}. 

\subsubsection{$B_s\to\ell_1\ell_2$}

The branching fraction for the mode $B_s\to\ell_1\ell_2$ reads
\begin{align}\label{expr1}
\mathcal{B}(B_s\to &\ell_1^-\ell_2^+)^\mathrm{th} = \frac{\tau_{B_s}}{64\pi^3}\frac{\alpha_\mathrm{em}^2 G_F^2}{m_{B_s}^3}f_{B_s}^3 |V_{tb}V_{ts}^\ast|^2 \lambda^{1/2}(m_{B_s},m_1,m_2)\nonumber\\
&\times\Bigg{\lbrace}[m_{B_s}^2-(m_1+m_2)^2]\cdot \Bigg{|}(C_9-C_9^\prime)(m_1-m_2)+(C_S-C_S^\prime)\frac{m_{B_s}^2}{m_b+m_s} \Bigg{|}^2 \nonumber \\
&\quad+[m_{B_s}^2-(m_1-m_2)^2]\cdot \Bigg{|}(C_{10}-C_{10}^\prime)(m_1+m_2)+(C_P-C_P^\prime)\frac{m_{B_s}^2}{m_b+m_s} \Bigg{|}^2 \Bigg{\rbrace},
\end{align}
where the $B_s$-meson decay constant $f_{B_s}$ is defined by $\Braket{
  0 |\bar{b} \gamma_\mu \gamma_5 s | B_s(p) } = i p_\mu f_{B_s}$.

\subsubsection{$B\to K\ell_1\ell_2$}

The differential branching ratio of $\bar{B}\to \bar{K}\ell_1^- \ell_2^+$ is given by
\begin{align}\label{expr2}
\dfrac{\mathrm{d}\mathcal{B}}{\mathrm{d}q^2}(\bar{B}\to \bar{K}\ell_1^- \ell_2^+) = & \left\vert\mathcal{N}_K(q^2)\right\vert ^2 \times \Big{\lbrace} \varphi_7(q^2)|C_7+C_7^\prime|^2+\varphi_9(q^2)|C_9+C_9^\prime|^2+\varphi_{10}(q^2)|C_{10}+C_{10}^\prime|^2\nonumber \\
&+\varphi_{S}(q^2)|C_{S}+C_{S}^\prime|^2+\varphi_{P}(q^2)|C_{P}+C_{P}^\prime|^2+\varphi_{79}(q^2)\mathrm{Re}[(C_7+C_7^\prime) (C_9+C_9^\prime)^\ast]\nonumber \\
&+\varphi_{9S}(q^2)\mathrm{Re}[(C_9+C_9^\prime) (C_S+C_S^\prime)^\ast]+\varphi_{10P}(q^2)\mathrm{Re}[(C_{10}+C_{10}^\prime) (C_P+C_P^\prime)^\ast]\Big{\rbrace},
\end{align}
where $(m_1+m_2)^2\leq q^2\leq(m_B-m_K)^2$, and the explicit expressions for the phase-space functions $\varphi_i(q^2)$ write
\begin{align}
\varphi_{7}(q^2) &=  \frac{2 m_b^2|f_T(q^2)|^2}{(m_B+m_K)^2} \lambda_B\left[1-\frac{(m_1-m_2)^2}{q^2}-\frac{\lambda_q}{3 q^4}\right], \nonumber
\end{align} 
\begin{align}
\label{eq:C910coeff}
\begin{split}
%\varphi_{7}(q^2) &=  \frac{2 m_b^2|f_T(q^2)|^2}{(m_B+m_K)^2} \lambda_B\left[1-\frac{(m_1-m_2)^2}{q^2}-\frac{\lambda_q}{3 q^4}\right], \\[0.6em]
\varphi_{9(10)}(q^2)&=\frac{1}{2}|f_0(q^2)|^2(m_1\mp m_2)^2 \frac{(m_B^2-m_K^2)^2}{q^2} \left[1-\frac{(m_1\pm m_2)^2}{q^2}\right]  \\
&+\frac{1}{2}|f_+(q^2)|^2 \lambda_B\left[1-\frac{(m_1\mp m_2)^2}{q^2}-\frac{\lambda_q}{3 q^4}\right]  ,\\[0.6em]
\varphi_{79}(q^2)&=\frac{2 m_b f_+(q^2)f_T(q^2)}{m_B+m_K} \lambda_B\left[ 1-\frac{(m_1-m_2)^2}{q^2}-\frac{\lambda_q}{3 q^4}\right],\\[0.6em]
\varphi_{S (P)}(q^2)&=\frac{q^2 |f_0(q^2)|^2}{2(m_b-m_s)^2}(m_B^2-m_K^2)^2 \left[1-\frac{(m_1\pm m_2)^2}{q^2}\right], \\[0.6em]
\varphi_{10P (9S)}(q^2)&=\frac{|f_0(q^2)|^2}{m_b-m_s}(m_1\pm m_2)(m_B^2-m_K^2)^2\left[1-\frac{(m_1 \mp m_2)^2}{q^2}\right],\end{split}
\end{align}
with $\lambda_q \equiv \lambda(\sqrt{q^2},m_1,m_2)$, $\lambda_B \equiv \lambda(\sqrt{q^2},m_B,m_K)$, and
\begin{equation}
\left\vert \mathcal{N_K}(q^2)\right\vert^2=\tau_{B_d}\dfrac{\alpha_\mathrm{em}^2G_F^2|V_{tb}V_{ts}^\ast|^2}{512\pi^5 m_B^3}\frac{\lambda_q^{1/2}\lambda_B^{1/2}}{q^2}.
\end{equation}
These expressions are computed by using the (standard) parametrization of the hadronic matrix elements in terms of form factors $f_{+,0,T}(q^2)$, namely, 
\begin{align}
\langle \bar{K}(k)\vert \bar{s}\gamma_\mu b \vert \bar{B}(p) \rangle &= \left[ (p+k)_\mu-\frac{m_B^2-m_K^2}{q^2}q_\mu\right] f_+(q^2)+ \dfrac{m_B^2-m_K^2}{q^2} q_\mu f_0 (q^2),\\
\langle \bar{K}(k)\vert \bar{s}\sigma_{\mu\nu} b \vert \bar{B}(p) \rangle &= - i (p_\mu k_\nu - p_\nu k_\mu)\dfrac{2 f_T(q^2,\mu)}{m_B+m_K}.
\end{align}
\subsubsection{$B\to K^\ast\ell_1\ell_2$}

Finally, the differential branching ratio of $B\to \bar{K}^\ast \ell_1^- \ell_2^+$ is given by
\begin{equation}\label{expr3}
\dfrac{\mathrm{d}\mathcal{B}}{\mathrm{d}q^2} (B\to \bar{K}^\ast \to (K \pi) \ell_1^- \ell_2^+)= \dfrac{1}{4} \left[ 3 I_1^c(q^2)+6 I_1^s(q^2)-I_2^c(q^2)-2 I_2^s(q^2) \right], 
\end{equation}
where the relevant angular coefficients $I_i(q^2)$ read\footnote{The complete angular distribution can be found in~\cite{Becirevic:2016zri,Gratrex:2015hna}.}
\begin{align}
I_1^s(q^2) &= \Big{[} |A_\perp^L|^2 +|A_\parallel|^2+(L\to R)\Big{]} \dfrac{\lambda_q+2[q^4-(m_1^2-m_2^2)^2]}{4 q^4}+\dfrac{4 m_1 m_2}{q^2}\mathrm{Re}\left(A_\parallel^L A_\parallel^{R\ast}+A_\perp^L A_\perp^{R\ast}\right),\nonumber \\
I_1^c(q^2) &= \left[ |A_0^L|^2+|A_0^R|^2\right]\dfrac{q^4-(m_1^2-m_2^2)^2}{q^4}+\dfrac{8 m_1 m_2}{q^2} \mathrm{Re}\left( A_0^L A_0^{R\ast} - A_t^L A_t^{R\ast}\right)\nonumber\\
&\qquad \qquad\qquad\qquad-2\dfrac{(m_1^2-m_2^2)^2-q^2(m_1^2+m_2^2)}{q^4}\left(|A_t^L|^2+|A_t^R|^2\right),\\
I_2^s(q^2) &= \dfrac{\lambda_q}{4 q^4}[|A_\perp^L|^2+|A_\parallel|^2+(L\to R)],\nonumber\\
I_2^c(q^2) &= -\dfrac{\lambda_q}{q^4}(|A_0^L|^2+|A_0^R|^2),\nonumber
\end{align}
and the helicity amplitudes are 
\begin{align}
A_\perp^{L(R)} &= \mathcal{N}_{K^\ast}\sqrt{2} \lambda_B^{1/2}\left[[(C_9+C_9^\prime)\mp (C_{10}+C_{10}^\prime)]\dfrac{V(q^2)}{m_B+m_{K^\ast}} + \dfrac{2 m_b}{q^2}(C_7+C_7^\prime) T_1(q^2)\right],\\
A_\parallel^{L(R)} &=-\mathcal{N}_{K^\ast}\sqrt{2}(m_B^2-m_{K^\ast}^2)\left[\left[ (C_9-C_9^\prime)\mp (C_{10}-C_{10}^\prime)\right]\dfrac{A_1(q^2)}{m_B-m_{K^\ast}}+\dfrac{2 m_b}{q^2}(C_7-C_7^\prime) T_2(q^2)\right],\nonumber\\
A_0^{L(R)} &= -\dfrac{\mathcal{N}_{K^\ast}}{2 m_{K^\ast} \sqrt{q^2}}\left\lbrace2 m_b(C_7-C_7^\prime)\left[ (m_B^2+3 m_{K^\ast}^2-q^2)T_2(q^2)-\dfrac{\lambda_B T_3(q^2)}{m_B^2-m_{K^\ast}^2} \right]\right.\nonumber\\
&\left.+\left[(C_9-C_9^\prime)\mp(C_{10}-C_{10}^\prime)\right]\cdot\left[ (m_B^2-m_{K^\ast}^2-q^2)(m_B+m_{K^\ast})A_1(q^2)-\dfrac{\lambda_B A_2 (q^2)}{m_B+m_{K^\ast}}\right] \right\rbrace, \nonumber \\
A_t^{L(R)} &= - \mathcal{N}_{K^\ast}\dfrac{\lambda_B^{1/2}}{\sqrt{q^2}} \left[(C_9-C_9^\prime)\mp (C_{10}-C_{10}^\prime)+\dfrac{q^2}{m_b+m_s}\left( \dfrac{C_S-C_S^\prime}{m_1-m_2}\mp \dfrac{C_P-C_P^\prime}{m_1+m_2}\right) \right]A_0(q^2)\nonumber
\end{align}
with
\begin{equation}
\mathcal{N}_{K^\ast}(q^2)=V_{tb}V_{ts}^\ast \left[ \tau_{B_d}\frac{ \alpha_\mathrm{em}^2 G_F^2}{3\times 2^{10}\pi^5 m_B^3} \lambda_B^{1/2}\lambda_q^{1/2}\right]^{1/2}.
\end{equation}
In this case there are seven independent form-factors $V(q^2)$, $T_{1,2}(q^2)$, and $A_{0,1,2}(q^2)$ defined by
\begin{align}
\langle \bar{K}^\ast(k)|\bar{s}\gamma^\mu(1-\gamma_5) b|\bar{B}(p)\rangle &= \varepsilon_{\mu\nu\rho\sigma}\varepsilon^{\ast\nu}p^\rho k^\sigma \frac{2 V(q^2)}{m_B+m_{K^\ast}}-i \varepsilon_\mu^\ast(m_B+m_{K^\ast})A_1(q^2)\\[.3em] 
&+i(p+k)_\mu (\varepsilon^\ast \cdot q)\frac{A_2(q^2)}{m_B+m_{K^\ast}}+i q_\mu(\varepsilon^\ast \cdot q) \frac{2 m_{K^\ast}}{q^2}[A_3(q^2)-A_0(q^2)],\nonumber \\[.7em] 
\langle  \bar{K}^\ast(k)|\bar{s}\sigma_{\mu\nu}  q^\nu(1-\gamma_5) b|\bar{B}(p)\rangle &= 2 i  \varepsilon_{\mu\nu\rho\sigma}  \varepsilon^{\ast\nu}p^\rho k^\sigma T_1(q^2)  +   \left[\varepsilon_\mu^\ast(m_B^2-m_{K^\ast}^2)-(\varepsilon^\ast \cdot q)(2p-q)_\mu \right]T_2(q^2) \nonumber\\[.3em] 
&  +    (\varepsilon^\ast \cdot q)   \left[q_\mu - \frac{q^2}{m_B^2-m_{K^\ast}^2}(p+k)_\mu \right]T_3(q^2),
\end{align}
where $\varepsilon_\mu$ is the $K^\ast$ meson polarization vector, and $q=p-k$. The form factor $A_3(q^2)$ is related to the others as $2 m_{K^\ast} A_3(q^2) = (m_B+m_{K^\ast})A_1(q^2)-(m_B-m_{K\ast})A_2(q^2)$.

\subsection{Hadronic quantities}
\label{app:had}

In this section we summarized the hadronic parameters considered in our analysis. The decay constants obtained by simulations of QCD on the Lattice are summarized in Tab.~\ref{tab:hadcons} \cite{Aoki:2016frl}. In order to perform the fit of $\mathcal{B}(B\to K \mu^+\mu^-)_{\text{high}-q^2}$, we combine the $B\to K$ form factors which were precisely determined in the low-recoil region by Ref.~\cite{Bouchard:2013pna}, and more recently by Ref.~\cite{Bailey:2015dka}. Similarly, to compute $R_D$ we have used the $B\to D$ form factors recently computed in Ref.~\cite{Lattice:2015rga}. Finally, the LFV semileptonic decays $B\to K^{(\ast)}\ell_1\ell_2$ were analysed by employing the form factors of Ref.~\cite{Ball:2004ye}.

\begin{table}[h!]
\renewcommand{\arraystretch}{1.5}
\centering
\begin{tabular}{| c |c|}
\hline 
\quad Quantity  &\quad Value [MeV]  \\ \hline\hline
$f_{K}$	&	$155.6(4)$  \\  
$f_{D}$	&	$212(2)$ 	\\  
$f_{D_s}$	&	$249(1)$  \\  
$f_{B}$	&	$186(4)$ 	\\  
$f_{B_s}$	&	$224(5)$ 	\\ \hline
$f_{\phi}$	&	$241(18)$ 	\\ \hline
\end{tabular}
\caption{\label{tab:hadcons}\small \sl  Decay constants used in the analyses. All values are taken from Ref.~\cite{Aoki:2016frl}, 
except for $f_\phi$ which was recently computed in Ref.~\cite{cd2}.}
\end{table}

\end{appendix}

\end{document}